\documentclass[aps,prd,nofootinbib,twocolumn,showpacs,superscriptaddress,groupedaddress]{revtex4}  
\usepackage{graphicx}  
\usepackage{dcolumn}   
\usepackage{bm}        
\usepackage{amssymb}   
\usepackage{amsmath}
\usepackage{amsfonts}
\usepackage{caption}
\usepackage{gensymb}
\usepackage{listings}
\usepackage{pythontex}
\usepackage{makecell}

\usepackage{ulem} 

\begin{document}

\widetext
\leftline{Accepted
for publication in PRD}


\title{Signal estimation in On/Off measurements including event-by-event variables}

\author{G.~D'Amico} \email{Email: giacomo.damico@uib.no}
\affiliation{Department for Physics and Technology, University of Bergen, Bergen, Norway}

\author{\\T.~Terzić}
\affiliation{University of Rijeka, Department of Physics, 51000 Rijeka, Croatia}

\author{J.~Strišković}
\affiliation{Josip Juraj Strossmayer University of Osijek, Department of Physics, 31000 Osijek, Croatia}

\author{M.~Doro}
\affiliation{Universit\`{a} di Padova and INFN, I-35131 Padova, Italy}

\author{M.~Strzys}
\affiliation{Institute for Cosmic Ray Research (ICRR), The University of Tokyo, Kashiwa, 277-8582 Chiba, Japan}

\author{J. van Scherpenberg}
\affiliation{Max-Planck-Institut f\"ur Physik, D-80805 M\"unchen, Germany}

\begin{abstract}
Signal estimation in the presence of background noise is a common problem in several scientific disciplines. An ``On/Off" measurement is performed when the background itself is not known, being estimated from a background control sample. 
The ``frequentist'' and Bayesian approaches for signal estimation in On/Off measurements are reviewed and compared, focusing on the weakness of the former and on the advantages of the latter in correctly addressing the Poissonian nature of the problem.
In this work, we devise a novel reconstruction method, dubbed BASiL (Bayesian Analysis including  Single-event Likelihoods), for estimating the signal rate based on the Bayesian formalism. It uses information on event-by-event individual parameters and their distribution for the  signal and background population. Events are thereby weighted according to their likelihood of being a signal or a background event and background suppression can be achieved without performing fixed fiducial cuts.
Throughout the work, we maintain a general notation, that allows to apply the method generically, and provide a performance test using real data and simulations of observations with the MAGIC telescopes, as demonstration of the performance for Cherenkov telescopes. BASiL allows to estimate the signal more precisely, avoiding loss of exposure due to signal extraction cuts. We expect its applicability to be straightforward in similar cases.
\end{abstract}

\maketitle

\section{Introduction}
\label{Sec:Intro}

In some experiments, where besides the signal the background is unknown, the signal itself can be obtained by a so-called "On/Off" measurement: a background-control (Off) region, which is supposedly void of any signal, is defined to estimate the background rate $b$. 
The ``On source'' measurement instead provides an estimate of the signal rate $s$ plus $b$, with the latter term supposed to be equal to that in the Off region. 
A normalization factor $\alpha$ between the On and Off exposure is normally introduced. If, for instance, the On and Off regions have the same acceptance, then $\alpha$ is  defined as the ratio of the effective observation time in the two regions: $\alpha=t_{on}/t_{off}$\footnote{A more detailed definition and discussion of \(\alpha\) can be found in Ref. \cite{berge2007background}.}. 
The measurement of the number of events in the On and Off region results in independent positive count numbers $N_{on}$ and $N_{off}$. 
 If one exactly knew the signal flux $s$,  the number of signal events $N_s$ in the On measurement would be a random variable following  a  Poisson distribution\footnote{ The generic symbol $ p()  $ is used to indicate all probability density functions (PDFs) and probability mass functions (PMFs) (the former applies to continuous variables while the latter to discrete variables). Note that the order of arguments is irrelevant being $ p( x, y | I )  $ the ``joint PDF (or PMF) of $ x  $ and $ y $ under condition $ I  $''.}
\begin{equation}
p( N_s \; | \;  s   ) = \frac{ (\xi \cdot s )^{N_s} e^{-\xi \cdot s} }{N_s !} , \quad  s \geq 0.
\label{Eq:Poisson_Signal}
\end{equation}
Where
\begin{align*}
 \xi =  t_{eff} \; \cdot A_{eff}
\end{align*}
is the exposure, with $t_{eff}$ being the effective observation time and $ A_{eff}$ the effective area of the telescope. For simplicity of notation, throughout the paper we will assume $\xi = 1 $ and we will refer to $s$ and $b$ as the signal and background rate, respectively. A summary of the variables used and their description can be found in Tab. \ref{tab:Variables}.

\begin{table*}[t]
\caption{\label{tab:Variables} Summary of the variables  with their description considered in an On/Off measurement and used throughout the paper. 
}

\begin{ruledtabular}
\begin{tabular}{l|l|l| l}
      variable       & description                   & property  & probability distribution \\ \hline
     $N_{on}$        &  number of events in the On region     &  measured &          \\  
     $N_{off}$       &  number of events in the Off region    &  measured &          \\  
     $\alpha$       &   exposure in the On region over the one in the Off regions & measured & \\
     $b$            &  expected rate of  occurrences of background events in the Off regions                   & unknown &  Eq.~\eqref{Eq:General_Likel} in which $s$ is integrated out  \\
      $s$            &   expected rate of  occurrences of signal events in the On region                     & unknown & Eq.~\eqref{Eq:Signal_PDF} \\
      $N_s$       &  number of signal events in the On region    &  unknown      & Eq.~\eqref{Eq:PDF_Nexc}    
\end{tabular}
\end{ruledtabular}
\end{table*}

The difficulty in estimating the signal rate $s$ lies in the uncertainties connected to the determination of the number of signal events $N_s$ from the measured counts $N_{on}$ and $N_{off}$, especially in the case of small Signal to Noise Ratio (SNR).
It is also important to underline that, because of the Poissonian nature of the problem (see Eq.~\eqref{Eq:Poisson_Signal}), both $s$ and $N_s$ must be non negative.

Assuming flat priors $p(s)$ and $p(b)$ (with $s>0$ and $b>0$) and by applying the Bayes theorem, we get that the PDF for the signal rate $s$ is
\begin{align}
p( s & \; | \;  N_{on},  N_{off}; \alpha)   \nonumber \\
& =  \frac{\int db \;  p(  N_{on}, N_{off} \; | \;  s,b ; \alpha) p(b)\,p(s)  }{\int ds \; db \;  p(  N_{on}, N_{off}, s,b ; \alpha)   }  \nonumber \\
& \propto  \int db \;  p( N_{on}, N_{off} \; | \;  s,b ; \alpha).
\label{Eq:PDF_Likelihood}
\end{align}

Thus the PDF of the signal rate $s$ is proportional to the likelihood function in which the background rate $b$ is integrated out, leaving a marginal distribution of $s$. 

The likelihood function can be expressed in the following way:
\begin{align}
& p( N_{on}, N_{off} \; | \;  s,b ; \alpha) = \, p( N_{on}  \; | \;  s, \alpha b) \cdot p(  N_{off} \; | \;  b) \nonumber \\
 & =   \frac{(s + \alpha b)^{N_{on}}}{N_{on}!} e^{- (s +\alpha b) } \cdot \frac{ b^{N_{off}}}{N_{off}!} e^{-  b },
 \label{Eq:Likelihood}
\end{align}
where we have made use of the independence of the measured values $N_{on}$ and $N_{off}$ and of the fact that both values come from a Poisson process with rate given respectively by $s +\alpha b$ and $b$.

Using the binomial identity\footnote{For reasons that will be clear in a while, the bound variable in the binomial identity is called $N_s$, i.e.
$$
(s + \alpha	b)^{N_{on}} = \sum_{N_s=0}^{N_{on}} \frac{N_{on}!}{(N_{on} - N_s)! N_s!} \; s^{N_s} (\alpha b)^{N_{on}-N_s}.
\label{Eq:Identity}
$$
}, we can factorize the likelihood in Eq.~\eqref{Eq:Likelihood} in two Poisson distributions, one for $N_s$ with expected value $s$ and one for  $N_{on} + N_{off} - N_s$ with expected value $b(1+\alpha)$:

\begin{align}
& p( N_{on},   N_{off} \; | \;  s,b ;   \alpha) 
\propto \nonumber   \\ 
&  \sum_{N_s = 0}^{N_{on}}  \frac{ (N_{on} + N_{off} - N_s)! }{(1+1/\alpha)^{-N_s}  (N_{on} - N_s)! }   \cdot  \frac{s^{N_s}}{N_s !} e^{-s} \nonumber \\
& \times \, \frac{(b(1+\alpha))^{N_{on} + N_{off} - N_s} }{ (N_{on} + N_{off} - N_s)!} e^{-b(1+\alpha)}.
\label{Eq:General_Likel}
\end{align}
Here, factors that depend merely on $N_{on}$ and $N_{off}$ have been ignored.

The integral in Eq.~\eqref{Eq:PDF_Likelihood} is now straightforward 
\begin{align}
& p(  s \; | \;   N_{on}, N_{off}; \alpha)  \nonumber \\
& \propto  \sum_{N_s = 0}^{N_{on}}  \frac{ (N_{on} + N_{off} - N_s)! }{(1+1/\alpha)^{-N_s}  (N_{on} - N_s)! }   \cdot  \frac{s^{N_s}}{N_s !} e^{-s}.
\label{Eq:Signal_PDF}
\end{align}
Note that now $s$ and $N_s$ are both random variables, which can have only non negative values, in agreement with the Poissonian nature of the problem under study.

 We take into account the following identities:
  \begin{align}
p( s & \; | \;   N_{on}, N_{off}; \alpha) = \nonumber \\ 
& \sum_{N_s = 0}^{N_{on}} p(N_s \; | \; N_{on}, N_{off}; \alpha) \cdot p(s \; | \; N_s )
\label{Eq:SignaPDF}
\end{align}
and 
\begin{align}
p(s  \; | \; N_s )  = \xi \frac{(s\cdot \xi)^{N_s}}{N_s !} e^{-\xi \cdot s}.
\end{align}
The former results from marginalizing over the variable $N_s$. The latter is obtained from applying the Bayesian theorem with constant priors to the likelihood in Eq.~\eqref{Eq:Poisson_Signal}. Recalling that $\xi = 1$,  we can now compare  Eq.~\eqref{Eq:Signal_PDF} and Eq.~\eqref{Eq:SignaPDF} to obtain the PMF of the variable $N_s$
\begin{equation}
p(N_s \; | \; N_{on}, N_{off}; \alpha) \propto  \frac{ (N_{on} + N_{off} - N_s)! }{(1+1/\alpha)^{-N_s}  (N_{on} - N_s)! }\,.
\label{Eq:PDF_Nexc}
\end{equation}

Eq.~\eqref{Eq:PDF_Nexc} allows then to define the most probable value (the mode) as an estimation of the number of signal events. This procedure was in fact previously outlined by T.~J.~Loredo (see Eq.~(5.13) of Ref.~\cite{Loredo}) already in 1992. The main goal of this work is to extend Eq.~\eqref{Eq:PDF_Nexc} by including the information of the individual events without limiting ourselves with a ``global'' method that makes use only of the number $N_{on}$ and $N_{off}$. 
To do so, we will first use Monte Carlo (MC) simulations to compare the Bayesian approach (summarized in Eq.~\eqref{Eq:Signal_PDF}) with the frequentist approach in Sec.~\ref{Sec:Comparison}. We will also discuss why the former is preferable in this problem. 
Then in Sec.~\ref{Sec:Single-events-Obs}, we will explain how to introduce single event information in Eq.~\eqref{Eq:PDF_Nexc}, and in Sec.~\ref{Sec:IACT_simulations} we will investigate the effects on the precision in the estimation of the number of the signal rate,  using as an example real data and simulations from the MAGIC Imaging Atmospheric Cherenkov telescopes (IACTs).

\section{Comparison between frequentist and Bayesian approach}
\label{Sec:Comparison}
In the previous section, we estimated the signal rate using a Bayesian approach. In literature however, the On/Off measurement problem is often solved in the frequentist approach.

In the frequentist approach \cite{Li_Ma,Rolke,mattox1996likelihood}, the background rate $b$, that is a nuisance parameter in the Bayesian approach, is not integrated out as done in Eq.~\eqref{Eq:PDF_Likelihood}. Instead from the likelihood in Eq.~\eqref{Eq:Likelihood} one defines the following test statistic (usually referred to as the likelihood ratio) 
\begin{equation}
\lambda(s) \equiv  \frac{  p( N_{on}, N_{off} \; | \;  s,b = \hat{b} \;  ; \;  \alpha) }{ p( N_{on}, N_{off} \; | \;  s = N_{on} - \alpha N_{off} \; , \; b =N_{off}  \; ;  \; \alpha) }, 
\label{Eq:LikeRatio}
\end{equation}
where\footnote{Note that when the null hypothesis is assumed ($s=0$), then $N = N_{on} + N_{off} $, $\hat{b} = \alpha N/(1+ \alpha)$ and Eq.~\eqref{Eq:LikeRatio} gives the Eq. (17) of Ref. \cite{Li_Ma} for computing the detection significance.}
\begin{equation}
\hat{b} = \frac{N + \sqrt{N^2 + 4(1+ \alpha)s N_{off}} }{2(1+ \alpha)}.
\end{equation}

is the value of $b$ that maximizes the likelihood in~Eq.~\eqref{Eq:Likelihood} for a given $s$, and $N \equiv N_{on} + N_{off} - (1+ 1/\alpha)s$.

The advantage of Eq.~\eqref{Eq:LikeRatio} is that according to Wilks' theorem  \cite{Wilks}, the function $-2 \log \lambda(s)$  has an approximate $\chi^2$ distribution with
1 degree of freedom, which can be used to extract confidence intervals. For example if we want the $68\%$ confidence interval we impose $-2 \log \lambda(s) = 1$ and it can be shown~\cite{Li_Ma} that for a large number of counts $N_{on}$ and $N_{off}$ this condition is satisfied 
when\footnote{ 
The factor after ``$\pm$'' in Eq.~\eqref{Eq:EstimatedS_LM} is derived from the variance of the linear combination of independent random variables.}


\begin{equation}
s = (N_{on} - \alpha N_{off}) \pm \sqrt{N_{on} + \alpha^2 N_{off}}\, .
\label{Eq:EstimatedS_LM}
\end{equation}

By imposing $-2 \log \lambda(s) = 3.84$ one can get $95\%$ upper limits, as it is done in Ref. \cite{Rolke}, although with \textit{ad-hoc} adjustments. These \textit{ad-hoc} adjustments are not surprising because the maximum likelihood approach described so far suffers from the following problems \cite{Loredo}:
\begin{itemize}

\item It only works well with counts number large enough. It is not suitable for low count numbers \cite{Li_Ma}, while the Bayesian approach has no limitations on that.

\item Only information about confidence intervals can be extracted. Legitimate questions such as ``What is the probability of having $N_s$ signal events in a sample of $N_{on}$ events?" cannot be answered (this is indeed possible in the Bayesian approach as shown in Eq.~\eqref{Eq:PDF_Nexc}).

\item  The frequentist result of Eq.~\eqref{Eq:EstimatedS_LM} does not exclude negative rate, but a Poisson process conflicts with negative rate\footnote{One can argue that in the frequentist approach these negative rates are in the end put equal to zero and a negative flux will not be claimed. But while this comes naturally in the Bayesian approach, in the frequentist approach instead this needs to be done by ``hand''  with the introduction of \textit{ad-hoc} adjustments.}.

\end{itemize}

\begin{figure*}
  \includegraphics[width=0.38\linewidth]{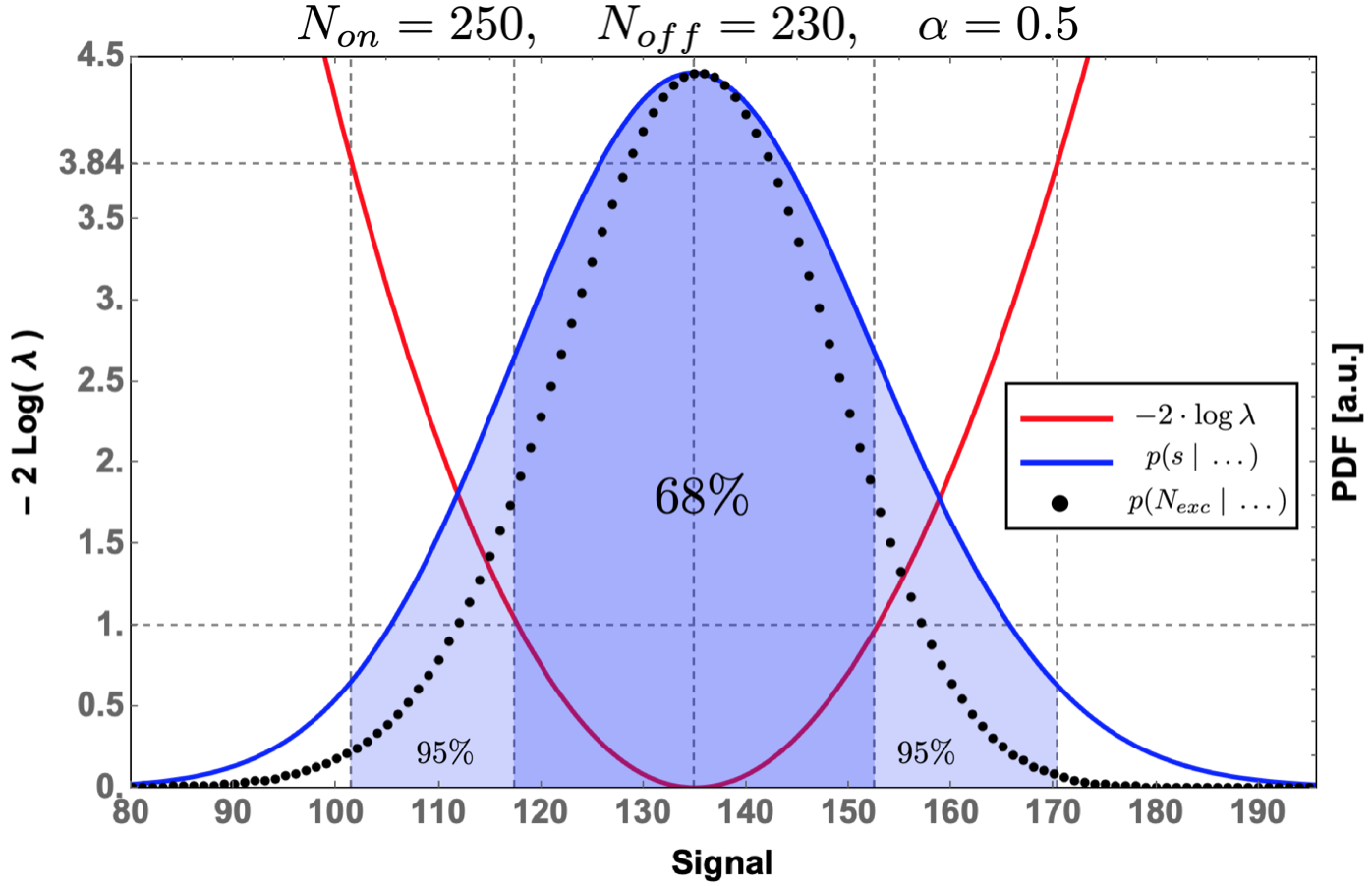}
  \includegraphics[width=0.38\linewidth]{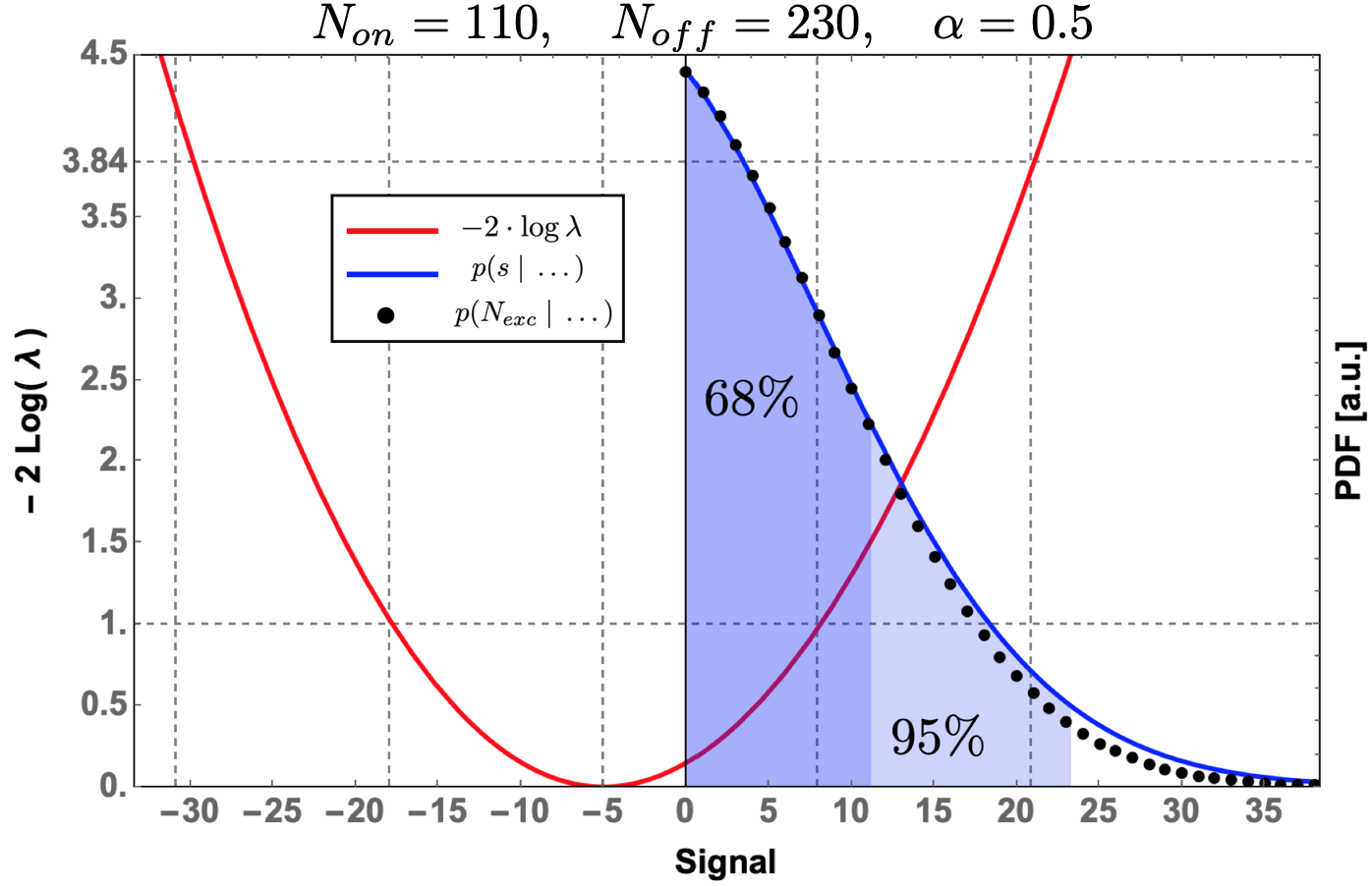}
\\
  \includegraphics[width=0.38\linewidth]{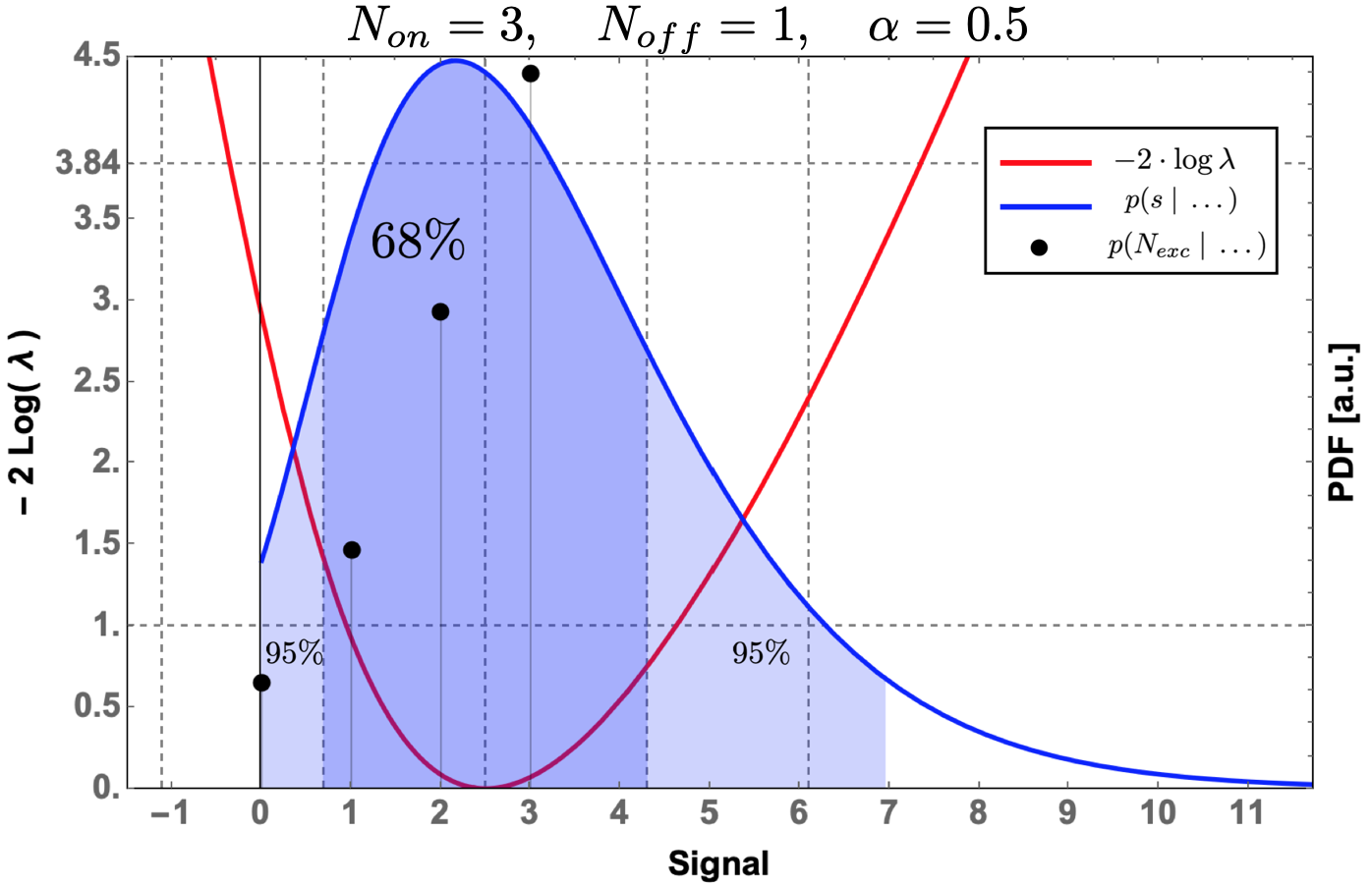}
  \includegraphics[width=0.38\linewidth]{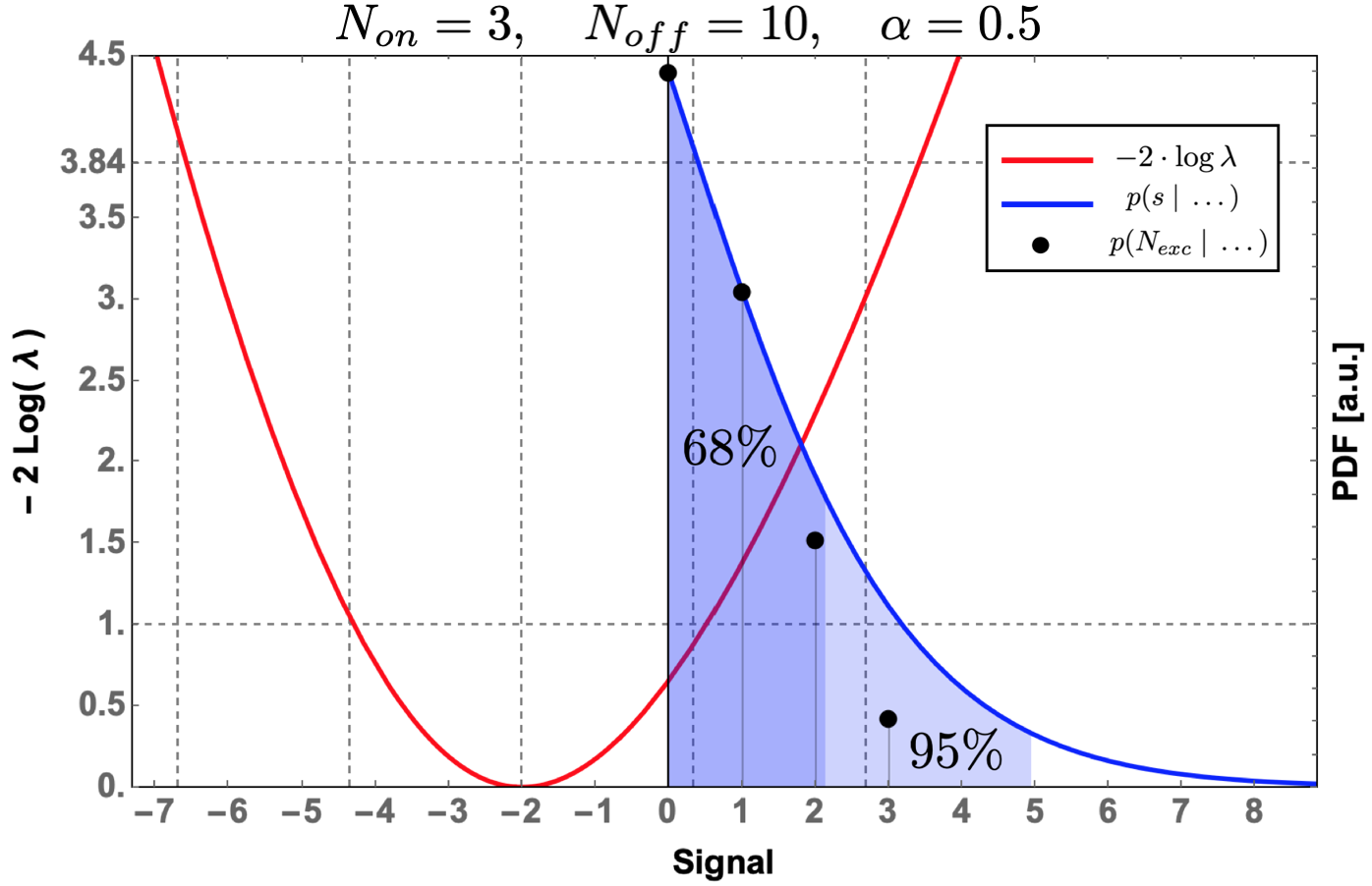}
\caption{Comparison for different values of $N_{on}$ and $N_{off}$ between $-2 \log \lambda(s)$ (red line) defined in Eq.~\eqref{Eq:LikeRatio} and $p( s \; | \;   N_{on}, N_{off}; \alpha) $ (blue line) and $p( N_s \; | \;   N_{on}, N_{off}; \alpha)$ (black points), defined respectively in Eq.~\eqref{Eq:Signal_PDF} and Eq.~\eqref{Eq:PDF_Nexc}. For the last distribution the x-axis does not show the signal rate but the discrete variable $N_s$. The last two probability distributions have been re-scaled for comparison with $-2 \log \lambda(s)$ . Vertical dashed lines are for  $s =E(s) \pm k \sigma_s $ ($k= \pm \{0,1,2 \}$), where   $ E(s) = N_{on} - \alpha N_{off}$ and $\sigma_s = \sqrt{N_{on} + \alpha^2 N_{off}}$. }
\label{Fig:Comparison_MaxLike_PDF}
\end{figure*}

To overcome the above issues \textit{ad-hoc} adjustments are required\footnote{The word ``adjustments" is present in Ref. \cite{Rolke} 7 times.}.
Another advantage of the Bayesian approach is that, once we have the PDF of the signal rate, all information are encoded in $p( s \; | \;   N_{on}, N_{off}; \alpha)$ defined in Eq.~\eqref{Eq:SignaPDF}. From this equation, we can obtain the mode that maximizes $p( s \; | \;   N_{on}, N_{off}; \alpha)$, i.e. the most probable value, or the $68\%$ credible interval\footnote{Not to be confused with the frequentist confidence interval.}  
$\left[ s_{left}, s_{right} \right] $ with $s_{left}$ and $s_{right} $ such that
    
\begin{align}
& \int_{s_{left}}^{s_{right}} p( s \; | \;   N_{on}, N_{off}; \alpha) ds = 0.68, \quad \text{with} 
\label{Eq:CredibleInterval} \\
&  p( s_{left} \; | \;   N_{on}, N_{off}; \alpha) = p( s_{right} \; | \;   N_{on}, N_{off}; \alpha) . \nonumber
\end{align}
If this last condition cannot be fulfilled\footnote{When dealing with low excess events the Bayesian credible intervals can be highly asymmetric around the estimated signal rate, as shown for instance in the right plots of Fig.~\ref{Fig:Comparison_MaxLike_PDF} where $s_{left} = 0$.} then $s_{left} = 0$ and upper limits (ULs) on the signal rate can be computed.
The $95\%$  UL $s_{95}$  can be  intuitively defined by

\begin{equation}
        \int_{0}^{s_{95}} p( s \; | \;   N_{on}, N_{off}; \alpha) ds = 0.95.
        \label{Eq:UL95}
\end{equation}

These definitions in the Bayesian formalism of credible interval and UL were already explored in the context of On/Off measurements in $\gamma$-ray astronomy by the author of Ref.~\cite{knoetig2014signal}. Although in the work in Ref.~\cite{knoetig2014signal} Jeffreys’s (and not constant)
priors\footnote{See for instance Ref.~\cite{d1998jeffreys} for a review of the problem regarding the choice of the priors.} were assumed.
    
In Fig.~\ref{Fig:Comparison_MaxLike_PDF} we show the comparison between the frequentist and Bayesian approach in estimating the signal rate. 
One can notice that  $-2 \log \lambda(s)$, defined in Eq.~\eqref{Eq:LikeRatio}, always has the minimum value at $ s= N_{on} - \alpha N_{off}$. 
Only for large number of counts (see upper plots of  Fig.~\ref{Fig:Comparison_MaxLike_PDF}) $-2 \log \lambda(s) = 1$ when $s = (N_{on} - \alpha N_{off}) \pm \sqrt{N_{on} + \alpha^2 N_{off}}$. It is also interesting to notice that confidence and credible intervals agree for large number of counts and when we are not close to the border $s$=0. For low count numbers and when we are close to the border of the parameter space the frequentist approach has the problems previously discussed for which one needs \textit{ad-hoc} adjustments.

 We ran MC simulations to compare the results obtained with the two approaches. 
 In each MC simulation $N_{on}$ is generated by the sum of two Poisson random numbers with expected count $s$ and $\alpha  b$, respectively. $N_{off}$ is instead generated from a Poisson distribution with expected count $b$. Once $N_{on}$ and $N_{off}$ are obtained the inferred signal and its uncertainty are computed according to Eq.~\eqref{Eq:EstimatedS_LM} for the frequentist approach. 
In the Bayesian approach instead, the estimated signal is given by the most probable value, with uncertainty corresponding to the $68\%$ credible interval defined in Eq.~\eqref{Eq:CredibleInterval}, i.e. $(s_{right} -s_{left})/2$.

\begin{figure*}
   \includegraphics[width=0.4\linewidth,height=0.25\textwidth]{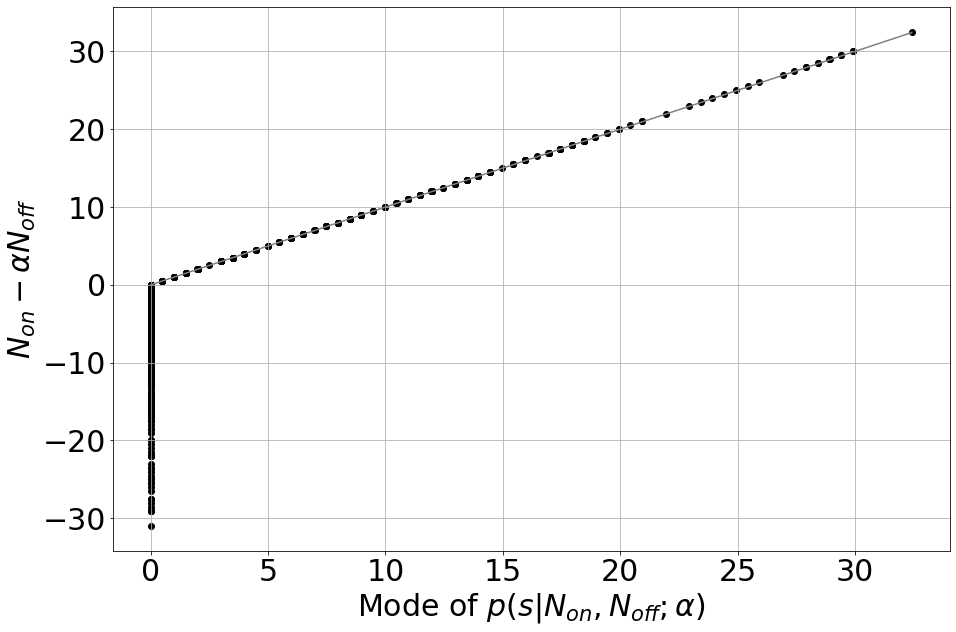}
   \includegraphics[width=0.4\linewidth,height=0.27\textwidth ]{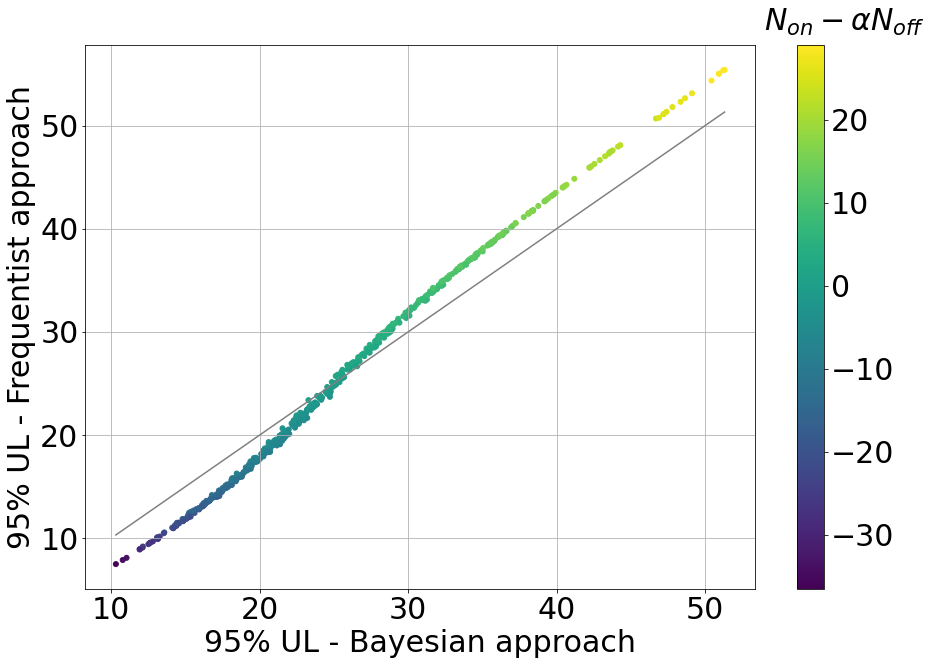}
\caption{Left: inferred signal from the Bayesian approach (x-axis) and  the frequentist approach (y-axis) using MC simulations. Right: 95 $\%$ UL obtained in the Bayesian approach (x-axis) from Eq.~\eqref{Eq:UL95} and in the frequentist approach (y-axis) following the prescription in Ref. \cite{Rolke} using MC simulations; color coding indicates the value of $N_{on}- \alpha N_{off}$. MC simulations were produced assuming $s=0$, $b=200$ and $\alpha = 0.5$.}
\label{Fig:Comparison_Approach}
\end{figure*}

Additionally, in the left plot of Fig.~\ref{Fig:Comparison_Approach} one can see that, as long as $N_{on} - \alpha N_{off} > 0$, there is a perfect agreement between the two approaches in estimating the signal rate. However, this is not anymore true  for $N_{on} - \alpha N_{off} < 0$.
In such case the Bayesian approach correctly (given the Poissonian nature of the problem) estimates a signal rate equal to zero. 
Similar considerations can be made for the $95\%$ UL estimation shown in the right plot of Fig.~\ref{Fig:Comparison_Approach}, where both approaches are in good agreement: as long as $N_{on} - \alpha N_{off} >0 $ a slight overestimation  of the UL value is observed for the frequentist approach \cite{Rolke} relative to the Bayesian one, while the opposite is true for $N_{on} - \alpha N_{off} <0$.

\begin{figure}
  \includegraphics[width=0.78\linewidth]{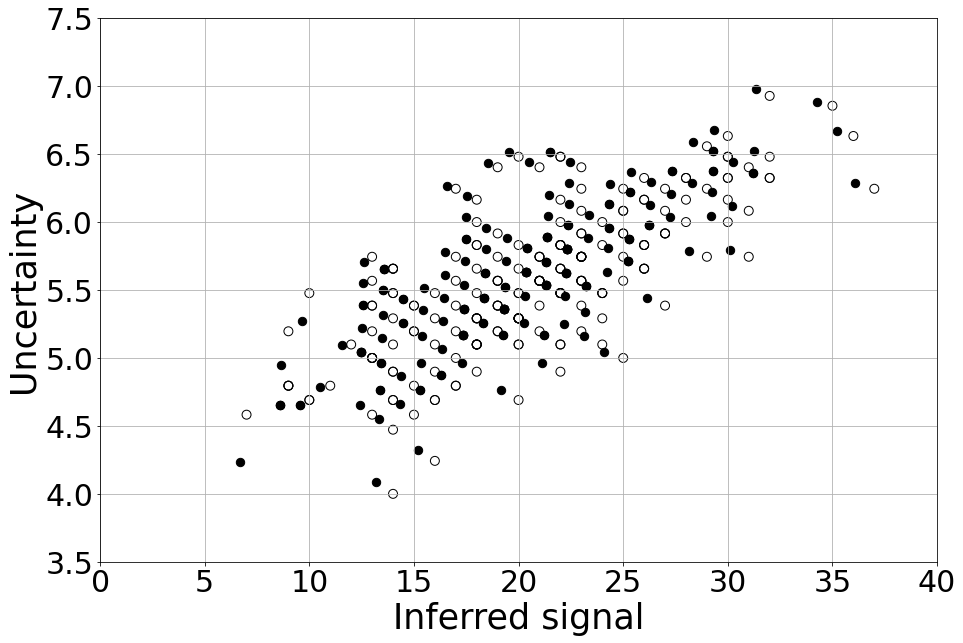}
\caption{Inferred signal (x-axis) and its uncertainty (y-axis) from MC simulations in which $s=20$, $b=5$ and $\alpha =1$. In the Bayesian approach (filled circles) the inferred signal is the mode of the signal PDF (see Eq.~\eqref{Eq:Signal_PDF}) with uncertainty given by $(s_{right}-s_{left})/2$ (see Eq.~\eqref{Eq:CredibleInterval}). In the frequentist approach (empty circles) the inferred signal is given by $N_{on} - \alpha N_{off}$ with uncertainty $\sqrt{N_{on} + \alpha^2 N_{off}}$.}
\label{Fig:Comparison_EstimatedSignal_Uncer}
\end{figure}

 In Fig.~\ref{Fig:Comparison_EstimatedSignal_Uncer} we show the $68\%$ confidence/credibility band (y-axis of the plot) around the estimated signal rate (x-axis of the plot) for both approaches, in which one can see that there is a good agreement between the results yielded by the frequentist and Bayesian approach.

\section{Probability density function of the signal rate including single-event observables}
\label{Sec:Single-events-Obs}
In Eq.~\eqref{Eq:PDF_Nexc}, we have defined the PMF of the number of signal events based on the number of events in the On and Off regions. It is common to select these events based on signal extraction cuts on one or more event variables to increase the SNR.
 A very common example of this in  astronomy is a cut performed in a region around the source, so that all events outside such region are ignored. A more advanced example is the implementation of some classification algorithm\footnote{This is the case of the gamma-hadron separation
for imaging Cherenkov telescopes, where each event is given a score
called \textit{Hadronness} in MAGIC \cite{MAGIC_RF}  or sometime referred to as \textit{Gammaness} in the Cherenkov Telescope Array (CTA) experiment.}  which yields for each event a discriminating variable that can be used for the background suppression. 

A disadvantage of cutting data is that also a fraction of the signal events will be excluded, which translates to a reduced exposure on the target. Moreover, normally after the selection, all events surviving a specific set of cuts are treated as equally probable signal (or background) events, regardless their ``distance'' from the cuts. We aim instead to fully exploit the information on how single-events variables distribute for a signal or a background population.
Our goal is to show how by replacing a fixed signal extraction cut with a statistical weighting of the events according to specific information (that is, not excluding any event a priori), we obtain a more precise signal estimation. We call this novel method Bayesian Analysis including  Single-event Likelihoods \texttt{BASiL}.

We start by including the information about the variables $\mathbf{x}$, which we have observed for each event, in the inference  of the signal rate $s$.
The variable $\mathbf{x}$ might be a single observable (like a discriminating variable obtained by a classification algorithm) or a set of observables.  Including $\vec{\mathbf{x}} = \{\mathbf{x}_1, \dots , \mathbf{x}_{N_{on}} \}$, Eq.~\eqref{Eq:PDF_Likelihood} becomes: 
\begin{equation}
p( s \, | \,  \vec{\mathbf{x}}, N_{on}, N_{off}; \alpha) \propto 
\int_0^{\infty} db \;  p( \vec{\mathbf{x}}, N_{on}, N_{off} \, | \,  s,b ; \alpha). 
\end{equation}

Using now the chain rule of probabilities, we write the likelihood in the following way:
\begin{align}
&p( \vec{\mathbf{x}}, N_{on}, N_{off} \; | \;  s,b ; \alpha) \nonumber \\
& =  p(\vec{\mathbf{x}} \; | \;  N_{on},   s, \alpha b) \cdot p(N_{on}  \; | \;    s, \alpha b) \cdot p(  N_{off} \; | \;  b).
\label{Eq:SignalLike_x_1}
\end{align}
The last two factors are Poisson distributions with expected counts $s + \alpha b$ and $b$, respectively.
The first factor is the probability of observing the variables $\vec{\mathbf{x}}$ in a sample of $N_{on}$ events with an assumed signal rate $s$ and background rate $\alpha b$. Given that all measured events are independent from each other, this probability is:
\begin{align}
& p(\vec{\mathbf{x}} \; | \;  N_{on},   s, \alpha b) \nonumber \\
&=  \prod_{i=1}^{N_{on}} \left[ \; p(\mathbf{x}_i \, | \, \gamma) \cdot p(\gamma \, | \, s, \, \alpha b  ) +  p(\mathbf{x}_i \, | \, \bar{\gamma}  ) \cdot  p(\bar{\gamma} \, | \,  s, \, \alpha b) 	\; \right],
\end{align}
where the term
\begin{align}
p(\gamma \, | \, s, \, \alpha b  )   = 1-   p(\bar{\gamma} \, | \,  s, \, \alpha b) =  \frac{s}{s+\alpha b}
\end{align}
is the \textit{prior} probability that one event is a signal event $\gamma$. We denote everything that is not signal as $\bar{\gamma}$. 

The terms $p(\mathbf{x} \, | \, \gamma)$ and $p(\mathbf{x} \, | \, \bar{\gamma}  ) $ are  the 
PDFs of observing the variables $\mathbf{x}$ from a signal or background population, respectively. In the Bayesian formalism, they can also be referred to as the likelihood functions of being a gamma or background event respectively, having observed the variables $\mathbf{x}$ for that particular event. 
Depending on the kind of variable or problem under study, these likelihoods can be estimated from MC simulations, a different data set or  be based on a theoretical model. One needs also to ensure that these likelihoods are normalized:
\begin{align}
\int_{\mathcal{X}} p(\mathbf{x} \, | \, \gamma) \, d\mathbf{x} = \int_{\mathcal{X}} p(\mathbf{x} \, | \, \bar{\gamma} )\,  d\mathbf{x} = 1,
\label{Eq:LikelihoodNormalization}
\end{align}
with $\mathcal{X}$ being the (multi-dimensional) parameter space in which the variables $\mathbf{x}$ are defined. 
The kind of likelihood in Eq.~\eqref{Eq:SignalLike_x_1} is known in literature \cite{conrad2015statistical}  as ``marked'' Poisson process, i.e. a Poisson process where each count is marked with a property (the variables $\mathbf{x}$ in our case) distributed according to a given PDF.

 Eq.~\eqref{Eq:SignalLike_x_1} can now be rewritten as (for a full derivation see Appendix~\ref{app:Derivation}):

\begin{align}
&  p( \vec{\mathbf{x}}, N_{on}, N_{off} \; | \;  s,b ; \alpha)   
\nonumber \\
& \propto \sum_{N_s = 0}^{N_{on}}  \frac{ (N_{on} + N_{off} - N_s)! }{(N_{on}  - N_s)! (1+1/\alpha)^{-N_s} } \;   \frac{ C(\vec{x}, N_s ) }{ \binom{N_{on}}{N_s}  } 
\nonumber \\
& \times \frac{s^{N_s}}{N_s !} e^{-s} \, \cdot \, \frac{(b(1+\alpha))^{N_{on} + N_{off} - N_s} }{ (N_{on} + N_{off} - N_s)!} e^{-b(1+\alpha)} \label{eq:LikelihoodH1_2}
\end{align}

where the function $C$ represents the combinatorial term:
\begin{align}
C(\vec{\mathbf{x}}, N_s ) = \sum_{A \in F_{N_s}} \prod_{i \in A}  p(\mathbf{x}_i | \gamma ) \cdot \prod_{j \in A^{c} }  p(\mathbf{x}_j | \bar{\gamma} )  
\label{Eq:combinatorialTerm}
\end{align}
with   $F_{N_s}$ being the set of all subsets of  $N_s$ integer numbers that can be selected from $\{1, \dots, N_{on}\}$. 

At this point, like done in Sec. \ref{Sec:Intro}, we can easily marginalize the nuisance parameter $b$ and obtain the final result for the PDF of the signal rate $s$:
\begin{align}
& p( s \; | \;  \vec{\mathbf{x}}, N_{on}, N_{off}; \alpha)  \propto \int db \;   p( \vec{\mathbf{x}}, N_{on}, N_{off} \; | \;  s,b ; \alpha) \nonumber  \\
& \propto \sum_{N_s = 0}^{N_{on}}  \frac{ (N_{on} + N_{off} - N_s)! }{(N_{on}  - N_s)!  (1+1/\alpha)^{-N_s}}    \frac{ C(\vec{\mathbf{x}}, N_s ) }{ \binom{N_{on}}{N_s}  } 
\nonumber \\
 & \times  \frac{s^{N_s}}{N_s !} e^{-s}.
\label{Eq:SignalPDFTotal}
\end{align}
Again one can recognize in this last expression the marginalization in Eq. \eqref{Eq:SignaPDF}, so that we can identify
\begin{align}
& p(N_s \; | \;  \vec{\mathbf{x}}, N_{on}, N_{off} ; \alpha) 
\nonumber \\
& \propto    \frac{ (N_{on} + N_{off} - N_s)! }{(N_{on}  - N_s)! (1+1/\alpha)^{-N_s} } \;    \frac{ C(\vec{\mathbf{x}}, N_s ) }{ \binom{N_{on}}{N_s}  }.
\label{Eq:Nexc_PDF_CombTerm}
\end{align}

 Given that the combinatorial term $C$, defined in Eq.~\eqref{Eq:combinatorialTerm}, is the novelty of this method, it is worth to elaborate  its role by providing an example. Let us assume $N_{on} = 3$ events in our On region and that we have also measured $x_1$, $x_2$, $x_3$ respectively for each event, with $x$ a variable whose distribution is $ p(x | \gamma ) $ for a signal population and $ p(x | \bar{\gamma} )$ for a background population. Thus, when $N_s = 0, 1, 2 ,3$ the combinatorial term will be respectively\footnote{In Appendix \ref{app:Algorithm}, a general algorithm is shown for efficiently   obtaining the term $C(\vec{\mathbf{x}}, N_s ) $, given the list of likelihoods $p(\mathbf{x}_i | \gamma ) $ and  $p(\mathbf{x}_j | \bar{\gamma} )  $ as input.}: 
 
\begin{align*}
C(\vec{x}, 0 ) = & p(x_1 | \bar{\gamma} ) \cdot p(x_2 | \bar{\gamma} ) \cdot p(x_3 | \bar{\gamma} ), \\
C(\vec{x}, 1 ) = &  p(x_1 | \gamma ) \cdot p(x_2 | \bar{\gamma} ) \cdot p(x_3 | \bar{\gamma} )  \\
& + p(x_1 | \bar{\gamma} ) \cdot p(x_2 | \gamma ) \cdot p(x_3 | \bar{\gamma} ) \\
& + p(x_1 | \bar{\gamma} ) \cdot p(x_2 | \bar{\gamma} ) \cdot p(x_3 | \gamma ), \\
C(\vec{x}, 2 ) = & p(x_1 | \gamma ) \cdot p(x_2 | \gamma ) \cdot p(x_3 | \bar{\gamma} )  \\
& + p(x_1 | \gamma ) \cdot p(x_2 | \bar{\gamma} ) \cdot p(x_3 | \gamma ) \\
 & + p(x_1 | \bar{\gamma} ) \cdot p(x_2 | \gamma ) \cdot p(x_3 | \gamma ), \\
 C(\vec{x}, 3 ) = & p(x_1 | \gamma ) \cdot p(x_2 | \gamma ) \cdot p(x_3 | \gamma ).
\end{align*}

From the above example it is clear how the combinatorial term is made up to account for all the possible combination of excess events among the total $N_{on}$ events that can give the observed values $\vec{\mathbf{x}}$. If, for instance
$$ p(\mathbf{x}_{i} | \gamma) = l \cdot p(\mathbf{x}_{i} | \bar{\gamma} ) \quad \forall i \in \{1,\dots, N_{on}\},
$$
i.e. all events are $l$ time more likely of being a signal event, then 
$$
 C(\vec{\mathbf{x}}, N_s )  \propto \binom{N_{on}}{N_s}  \; l^{N_s} 
$$
and 
\begin{align}
p(N_s & \; | \;  \vec{\mathbf{x}}, N_{on}, N_{off} ; \alpha) 
\nonumber \\
& \propto   \frac{ (N_{on} + N_{off} - N_s)! }{(N_{on}  - N_s)!  (1+1/\alpha)^{-N_s} } \cdot l^{N_s}.
\end{align}
By taking into account the information that all events are $l$ times more likely of being a signal event, we have updated the PMF of the number of signal events introducing a factor $l^{N_s}$. Its maximum values are obtained for $N_s = N_{on}$ if $l>1$, and for $N_s = 0$ if $l<1$. 
For $l=1$ we do not gain any information from the observed variable $x$, and we recover the result previously obtained in Sec.~\ref{Sec:Intro}.

With the introduction of the combinatorial term in Eq. \eqref{Eq:Nexc_PDF_CombTerm} we have devised the method to include event-by-event information for the computation of $N_s$. The power of this method clearly depends on the specifics of the datasets in which it is applied, and in turn, it depends on (i) the event parameters that are used, (ii) how they distribute for the signal and background population, and (iii) how performing is the signal extraction method that relies on a fixed fiducial cut. However, in order to be predictive and define a framework to assess the performance of the BASiL method, we apply it to a specific case, that of gamma-ray observation. For this purpose, we analyze real data from the  Major Atmospheric Gamma Imaging Cherenkov (MAGIC) Collaboration\footnote{\href{https://magic.mpp.mpg.de/}{https://magic.mpp.mpg.de/}}. 
Results reported in Ref. \citep{Performance_Paper} will be used as a benchmark case.

\section{The case of Imaging Atmospheric Cherenkov telescopes}
\label{Sec:IACT_simulations}

IACTs image the Cherenkov light emitted in the atmosphere by extended atmospheric showers generated by cosmic gamma rays (or cosmic rays) when entering the atmosphere. 
An irreducible background survives all possible image selection criteria and the signal estimation is performed through an ``On/Off'' comparison based, in which the Off sample is taken from a region in the sky where no signal is expected. 
 For steady point-like sources two variables are generally further used to suppress the background: the squared\footnote{Signal events spread around the region of interest and for a point-like source they distribute according to a 2-dimensional Gaussian distribution. Such a 2-dimensional Gaussian in the $\theta_{x}$ and $\theta_y$ space will correspond
to an exponential function for the distribution of $\theta^2 = \theta_{x}^2 + \theta_{y}^2 $.} angular distance from the source $\theta^2$, and 
a particle identification variable, which in the case of MAGIC is computed by means of a Random Forest (RF) algorithm, and is dubbed \textit{Hadronness} (h) \cite{MAGIC_RF}. 
The RF event classifier takes the image parameters of the event as input and returns a value between 0 and 1. The smaller the value, the more the event looks like a gamma-ray event. 
The $\theta^2$ parameter, related to the instrument point spread function depends on the telescope optics and mechanics, and mostly on the shower physics and image reconstruction (see Ref.~\cite{DaVela:2018ire}).
Therefore, the individual-event variables to consider are
$$
\mathbf{x} = (\theta^2, h, E)\,.
$$
Because the distributions of $\theta^2$ and $h$ are energy dependent, we have also included the estimated energy $E$\footnote{In principle one could also consider the time of arrival of individual events $t$ as element of $\mathbf{x}$. While this may be useful in some scenarios, $t$  can be neglected if we consider small enough time bins or similar conditions throughout the entire observation, that allows us to integrate out the time in our analysis.}.  

The likelihoods of being a signal $p( \mathbf{x} \, | \, \gamma )$ or a background $p( \mathbf{x} \, | \, \bar{\gamma})$ event can be factorized into three terms:
\begin{align}\label{eq:lkl}
& p( \mathbf{x} \, | \, \gamma ) = p( h \, | \, E, I, \gamma )\cdot p(  \theta^2 \, | \, E, I, \gamma ) \cdot p( E  \, | \,I, \gamma ), \nonumber \\
& p( \mathbf{x} \, | \, \bar{\gamma} ) =p( h \, | \, E, I, \bar{\gamma} )\cdot p(  \theta^2 \, | \, E, I, \bar{\gamma} ) \cdot p( E \, | \,I, \bar{\gamma} ).
\end{align}
where $I$ stands for the conditions under which the observation has been performed (e.g. zenith angle, atmospheric opacity etc.). 
 As the correlation between $\theta^2$ and $h$ range between zero and 0.2, approximately, we can consider the two variables as independent. 
The same is not valid for the correlation between $h, \theta^2$  and energy, which forces us to take into account the energy dependence of the distribution of $\theta^2$ and $h$, and apply the method in sufficiently small energy bins\footnote{If one wants to extrapolate the method to an unbinned analysis, a signal flux $p( E  \, | \,I, \gamma )$ has to be assumed beforehand.}.

We therefore focus on individual energy bins where the flux is assumed to be constant, so that $p( E  \, | \,I, \gamma )$ and $p( E  \, | \,I, \bar{\gamma} ) $ are uniform. This means that the factor  $p( E  \, | \,I, \gamma ) = p( E  \, | \,I, \bar{\gamma} ) $ is the same for all likelihoods in Eqs.~(\ref{eq:lkl}) and can be therefore ignored. Thus, to get the likelihood for each event of being a signal or background event one needs to only compute the distribution in $\theta^2$ and \textit{Hadronness} respectively from a signal and background population.

A sample of background events can be obtained by performing observations on regions of the sky (Off regions) where no signal contamination is observed and with similar conditions $I$ as the ``On'' sample, as explained in the Sec. \ref{Sec:Intro}. 
Obtaining a signal sample from the On region is less straightforward, because in the On region both a signal and an irreducible background contributions are present. For this reason in IACTs one has to rely on MC simulations of signal events to study the parameter distribution of a signal sample. Nonetheless, for a bright enough source like the Crab Nebula\footnote{The Crab Nebula is the brightest steady TeV gamma-ray emitter in the sky. It is a pulsar wind nebula which is used as a standard calibration for IACTs~\citep{Performance_Paper}.},  a very pure sample of  $\gamma$-ray signal events can be extracted from the On measurement, which allows us to study its properties.
Data are taken in the so-called wobble\footnote{In a wobble mode the source is placed with a certain offset with respect to the camera center during the observation. It allows simultaneous signal and background estimation.}
mode \cite{1994APh.....2..137F}, which yields $N_{on}$ and $N_{off}$ counts, then the excess is obtained by subtracting from the $N_{on}$ counts the (relative small) background count $\alpha N_{off}$ and the procedure is repeated for different cuts in \textit{Hadronness} or $\theta^2$. 

\begin{figure*}[h!t]
  \includegraphics[width=0.4\linewidth,height=0.25\textwidth]{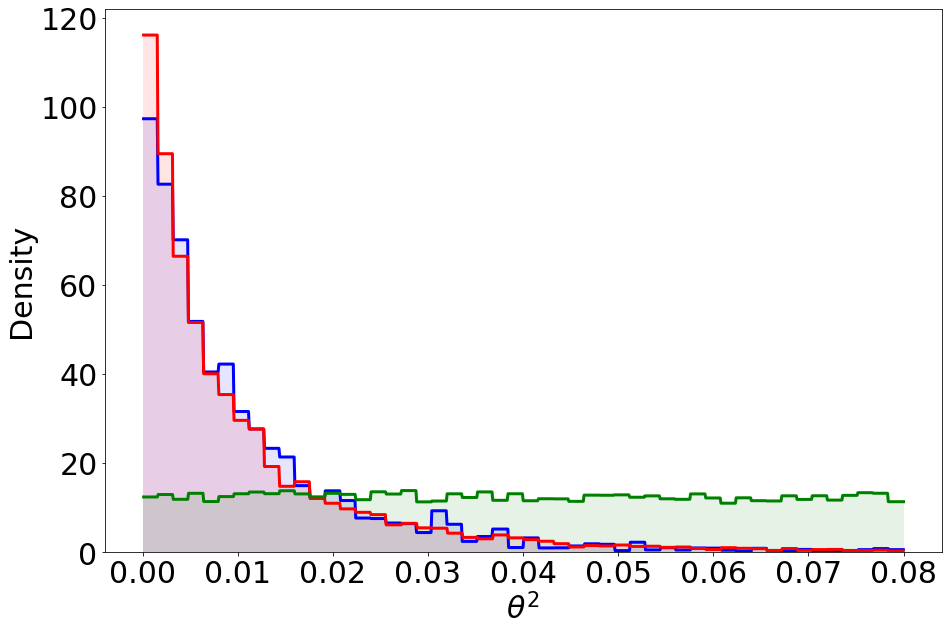} 
  \includegraphics[width=0.4\linewidth,height=0.25\textwidth]{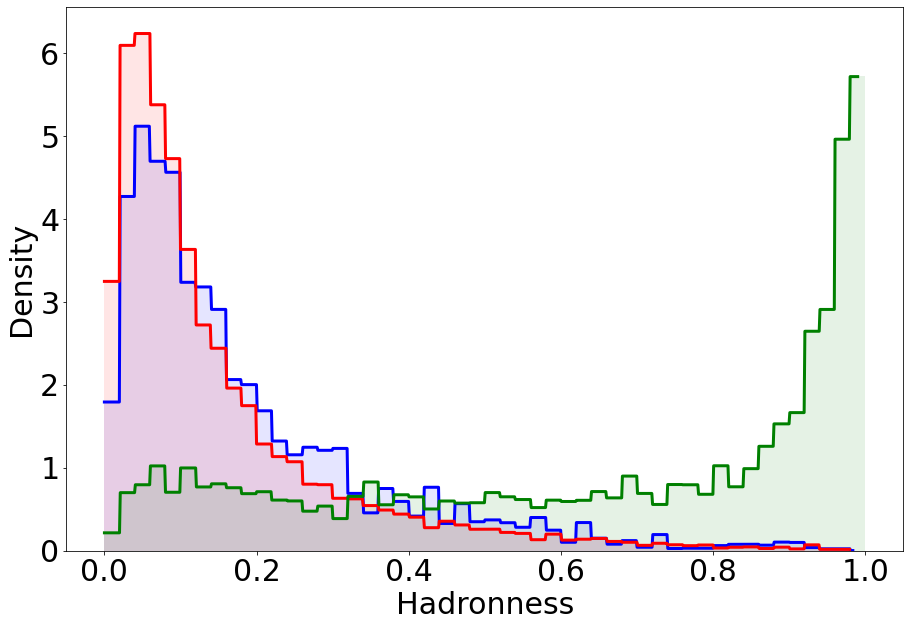}
  
\caption{Distribution in $\theta^2$ (left) and \textit{Hadronness} (right) for simulated (red) and observed (blue) signal excess and for background events (green) in the energy range 189-300 GeV. These distributions are obtained from Fig. 1 and 15 of Ref.~\cite{Performance_Paper}. For all histograms $10^5$ events were generated and divided in 50 bins. All distributions (colored areas in the figure) are normalized to 1. For a discussion on the origin and effect of the MC/data discrepancies see Ref.~\cite{Performance_Paper}.}
\label{Fig:Lkl_theta2_h}
\end{figure*}

Fig.~\ref{Fig:Lkl_theta2_h} shows the distribution in $\mathbf{x}=(h,\theta^2)$ of the signal excess from the Crab Nebula sample, MC-simulated signal and background events. For brevity, we show only events with estimated energy between 189 and 300 GeV.
Following the same prescription and using the same data set, a similar analysis is performed for $\theta^2$ which yields the distributions in the left plot of Fig.~\ref{Fig:Lkl_theta2_h}. 

A few important facts appear from the plots in Fig.~\ref{Fig:Lkl_theta2_h}:  (i) there is a mismatch between Monte Carlo data and real data, (ii) such difference is larger in \textit{Hadronness} than $\theta^2$, especially at very low \textit{Hadronness} values, where several signal events are not classified as gammas with sufficient degree of confidence,
(iii)  the signal selection based on $\theta^2$ is efficient with a cut at about 0.02~deg, that entails 75\% of the signal, while an optimal cut in \textit{Hadronness} is more complex to define, because it depends more strongly on the energy.

The optimization of the SNR can be done in several ways. The MAGIC collaboration elaborated a set of cuts specific for each energy bin, according to an ``efficiency'' parameter $\epsilon$ defined as the fraction of Monte Carlo signal events surviving a certain cut. 
In the following, we elaborate on this, and compare the outcome with the novel method which we propose.

Assuming a signal rate $s$ and background rate $b$, MAGIC observations are simulated following these steps:
\begin{itemize}
    \item generate $N_s$ and $N_{bkg}$  from a Poisson distribution with expected value  respectively $s$ and $\alpha \cdot b$, and define the number of counts in the On region as 
   \begin{align*}
         N_{on} = N_s + N_{bkg},
   \end{align*}
    \item  generate $N_{off}$, the total number of events in the Off region, from a Poisson distribution with expected value $b$,
    \item generate $\theta^2$ values for the events in the On region by randomly picking up $N_s$ values from the signal distributions (blue histogram in the left plot of Fig.~\ref{Fig:Lkl_theta2_h}) and $N_{bkg}$ values from the background distributions (green histogram in the left plot of Fig.~\ref{Fig:Lkl_theta2_h}),
    \item generate $\theta^2$ values for the events in the Off region by randomly picking up $N_{off}$ values from the background distributions (green histogram in the left plot of Fig.~\ref{Fig:Lkl_theta2_h})
    \item finally, the same is done for generating \textit{Hadronness} values for the On and Off measurements using this time the right plot of Fig.~\ref{Fig:Lkl_theta2_h}.
\end{itemize}

Having an On and Off measurement, we get an estimation $\hat{s}$ of the signal rate using only the information about the total counts $N_{on}$ and $N_{off}$ and the single-event variables $\mathbf{x}=(h,\theta^2)$.
This estimation is done using two different approaches, referred to as the ``standard'' and ``BASiL'' approach:
\begin{enumerate}

    \item The estimated signal rate is obtained from
    \begin{align*}
    \hat{s} = N_{on} - \alpha N_{off},
    \end{align*}
    where $N_{on}$ and $N_{off}$ are the numbers of events surviving the cut in $\theta^2$ and/or \textit{Hadronness} for the On and Off measurement, respectively. 
    Cut values are obtained assuming a given $\gamma$-ray efficiency $\epsilon$ computed from the signal distributions (see blue histograms of  Fig.~\ref{Fig:Lkl_theta2_h}). Being the most common way of suppressing the background and estimating $s$, we will refer to this approach as the ``standard'' one. 
    \item In the BASiL approach $\hat{s}$ is instead defined from the mode of the PMF\footnote{One could have also used the signal-rate PDF defined in Eq~\eqref{Eq:SignalPDFTotal}, but such choice would not change the results since the distributions for $s$ and $N_s$ share the same mode (one is simply derived from the other by including Poisson statistics).} defined in Eq.~\eqref{Eq:Nexc_PDF_CombTerm}, where $\mathbf{x}$ can be either $\theta^2$ and \textit{Hadronness}, or only one of them. 
    The combinatorial term in Eq.~\eqref{Eq:combinatorialTerm} will be obtained using signal and background likelihood values from the signal distributions (blue histograms in Fig.~\ref{Fig:Lkl_theta2_h}) and background distribution (green histograms in Fig.~\ref{Fig:Lkl_theta2_h}).

\end{enumerate}

It is important to stress that in both approaches the values of $s$, $b$, $N_s$ and $N_{bkg}$ are not taken into account: only observed quantities (counts in the On and Off regions, $\theta^2$ and \textit{Hadronness})  are considered for signal rate estimation. 
More importantly, for this study we ignore the mismatch between the ``real''-$\gamma$ and the MC-$\gamma$ distributions (respectively the blue and red histograms of Fig. \ref{Fig:Lkl_theta2_h}). In Appendix \ref{app:MC_real_mismatch} the study on how such mismatch between the MC and real data affects the estimation is shown. The conclusion is that this mismatch induce a bias in the estimation that leads to underestimate the number of excess events.

\begin{figure}[h!t]
  \includegraphics[width=0.95\linewidth]{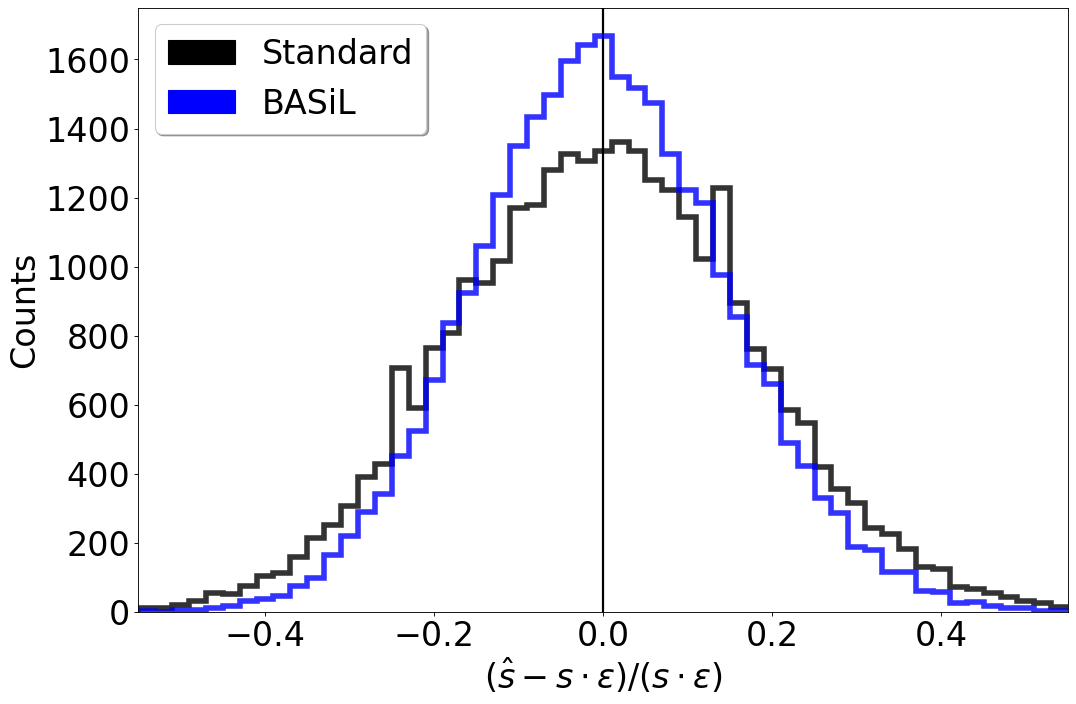} 
\caption{Distribution of $
( \hat{s} -s \cdot \epsilon )/(s\cdot \epsilon)$, 
with $\hat{s}$ the estimated signal rate obtained from $N_{on} - \alpha N_{off}$ in the standard approach (black), and from the mode of the distribution in Eq.~\eqref{Eq:Nexc_PDF_CombTerm} in the BASiL approach (blue). In the standard approach $N_{on}$ and $N_{off}$ are the number of events surviving the efficiency cut applied on the data. Efficiency is $95\%$ and $75\%$ respectively for \textit{Hadronness} and $\theta^2$, which translates in a total efficiency $ \epsilon = 67.6\%$ (see main text, Sec.~\ref{Sec:IACT_simulations} for details). 
In the BASiL approach no cut is applied on the data, i.e. $\epsilon = 1$. Values of $s$ and $b$ used in the simulations are $10^2$ and $3\cdot 10^3$, respectively, with $\alpha =1/3$.}
\label{Fig:Nexc_distribution}
\end{figure}

At this point we defined the signal-estimation precision and bias respectively as the standard deviation and mean value of the re-scaled distribution of $\hat{s}$, i.e. 
\begin{align}
     & prec. =  \sigma \left( \frac{\hat{s} - s\cdot \epsilon   }{s \cdot \epsilon} \right),  \label{Eq:Res_def}\\
    & bias = < \frac{\hat{s} - s\cdot \epsilon   }{s \cdot \epsilon} >. \label{Eq:Bias_def}
\end{align}
Note that the efficiency cut is $\epsilon=1$ in the BASiL approach, since no cut is applied on the data in this case. For a fair comparison in the standard approach $\hat{s}$ is put equal to zero whenever $N_{on} - \alpha N_{off} < 0$.  In Fig.~\ref{Fig:Nexc_distribution} an example of such distribution is shown,
where the signal rate is estimated using the standard (black histogram) and BASiL approach (blue histogram). 

\subsection{Signal-estimation precision and bias for different efficiency cut}
\label{Subsec:Res_bias_eff}

\begin{figure*}[h!t]
  \includegraphics[width=0.45\linewidth]{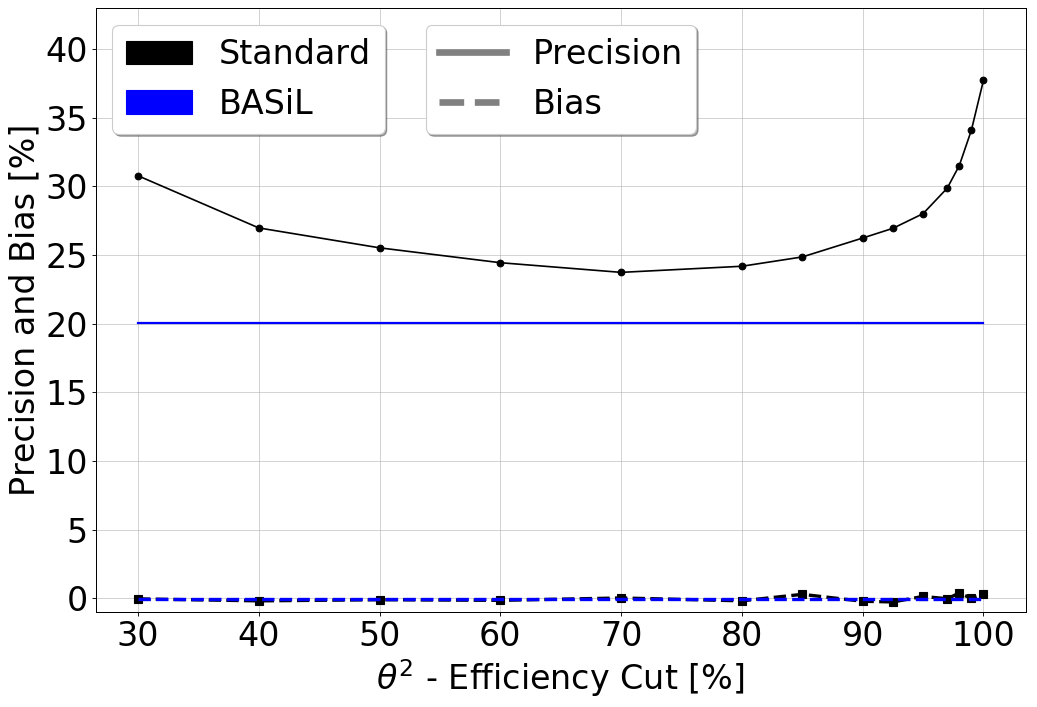} 
    \includegraphics[width=0.45\linewidth]{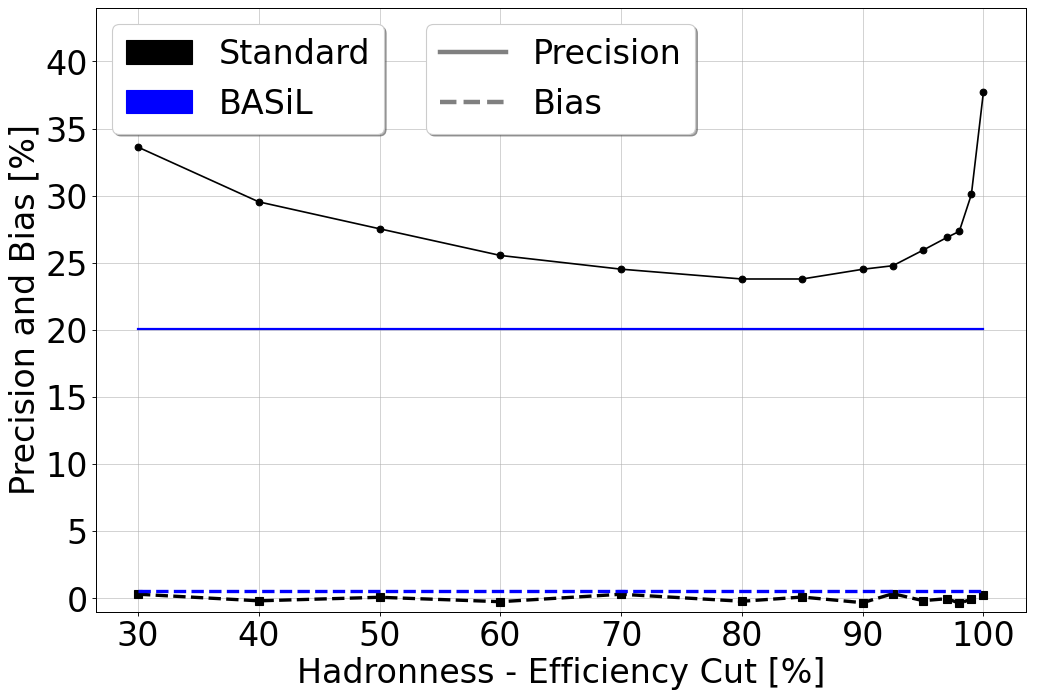} 
\caption{Comparison between the standard (black) and BASiL (blue) approach for the evolution of the precision (full line) and bias (dashed line) in the signal-rate estimation assuming different gamma-ray efficiency cut in $\theta^2$ (left) and \textit{Hadronness} (right). Note that for the BASiL approach the precision and bias do not depend on the efficiency, being $\epsilon =1$ in such case. Nonetheless, for a visual comparison these values are shown as horizontal lines. The definition of  precision and bias can be found in Eq.~\eqref{Eq:Res_def} and \eqref{Eq:Bias_def}, respectively. Observations are simulated assuming $s=10^2$ and $\alpha b = 10^3$, with $\alpha = 1/3$.   }
\label{Fig:ResBias_Eff}
\end{figure*}

We first study the evolution of the bias and precision defined in Eqs. (\ref{Eq:Res_def}) and (\ref{Eq:Bias_def}) for different efficiency cuts considering only $\theta^2$ or \textit{Hadronness} as single-event variable. 
For this study we assume a background intensity,  in the On region $\alpha b = 1000$ and a SNR of $10\%$, i.e. $s=100$. 
We then simulate observations following the steps previously described where in one case events have only $\theta^2$ as an observed variable and only \textit{Hadronness} in the other case.  Fig.~\ref{Fig:ResBias_Eff} reports the results. 

It is worth noticing that in the \textit{Hadronness} case, starting from 100\% efficiency, the precision in the standard approach improves immediately reaching its best value when the efficiency is about $80-90 \%$. 
    The same improvement happens also in the $\theta^2$ case, but it is smoother with a minimum around $70-80 \%$. An explanation of this effect can be easily found in Fig.~\ref{Fig:Lkl_theta2_h}, where it is clear that cutting in \textit{Hadronness} allows to suppress more background than performing a similar cut (i.e. with the same efficiency) in $\theta^2$. At low efficiency values the precision  is dominated by the Poisson statistic of the excess number and therefore its evolution follows 
    $$
    prec. \sim 1/\sqrt{N_s} \, \propto \,   1/\sqrt{\epsilon\;} \, .
    $$
    In the BASiL approach instead the precision  does not depend on the efficiency and it is about $15\%$ better than the best precision  we can achieve in the standard approach.
The bias in the signal estimation is very close to zero, apart from small fluctuations, in both approaches. As discussed in Appendix~\ref{app:MC_real_mismatch}, the most important source of bias is due to the non-perfect agreement between the real and simulated signal distribution. 

We therefore conclude that the BASiL method, by including the likelihood of each event of being a signal or background,  estimates the signal rate more precisely, while keeping the bias comparably low: for a SNR of $10\%$ the improvement in precision  is about $\sim 15\%$ in both \textit{Hadronness} and $\theta^2$.

\subsection{Signal-estimation precision  and bias for different signal to background ratio}

\begin{figure*}
  \includegraphics[width=0.45\linewidth]{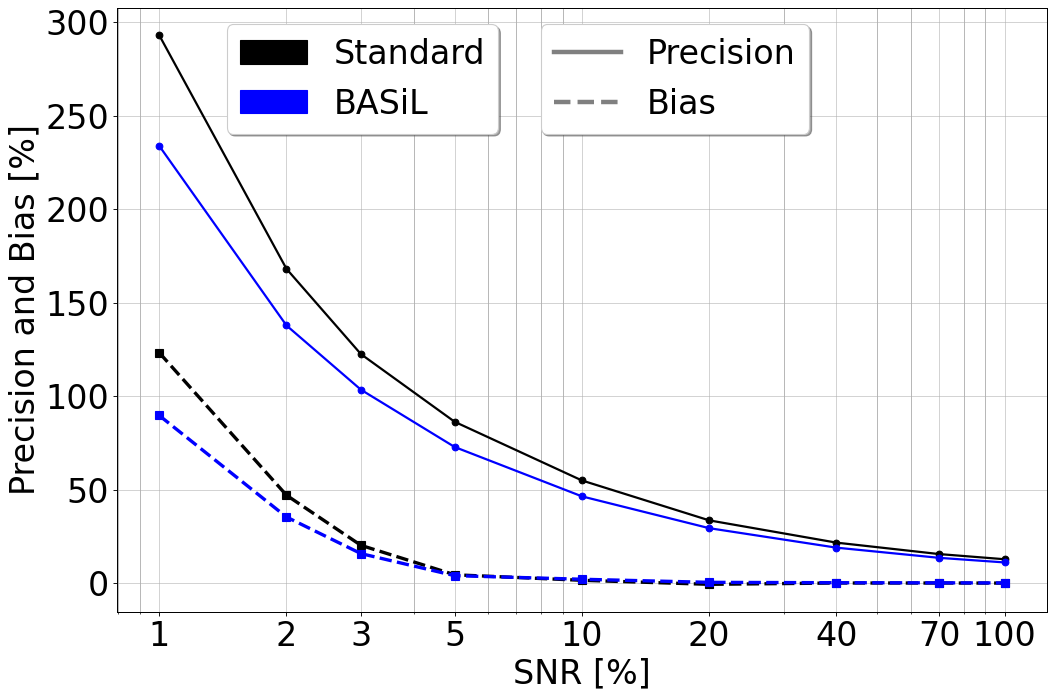} 
    \includegraphics[width=0.45\linewidth]{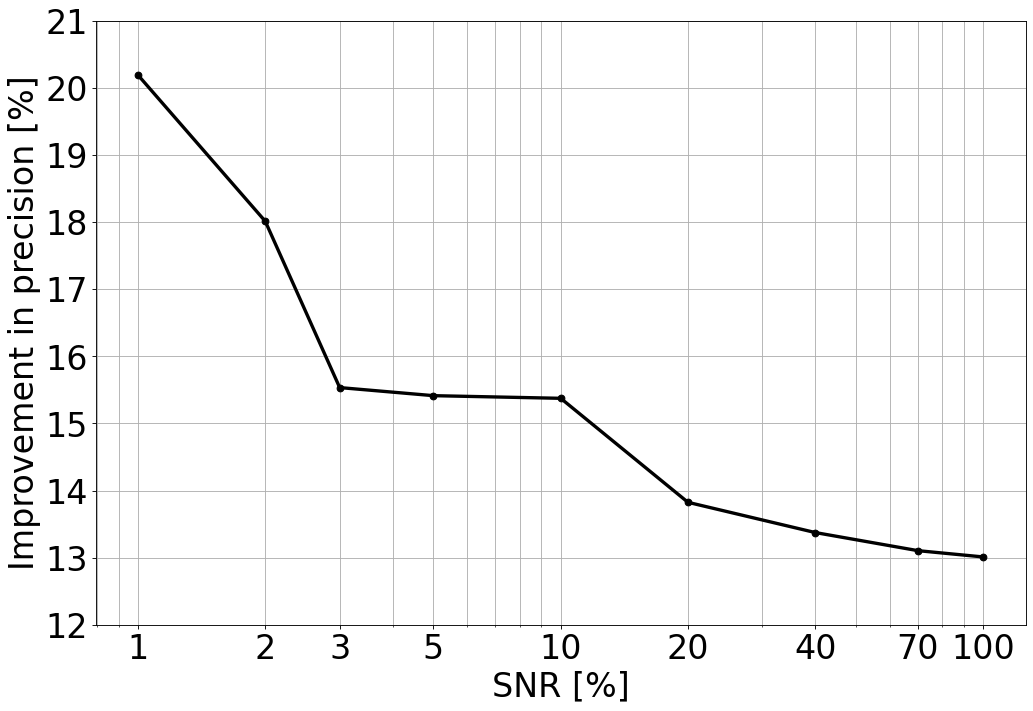} 
\caption{Left: comparison between the standard (black) and BASiL (blue) approach for the evolution of the precision  (full line) and bias (dashed line) assuming different SNR. In the standard approach the efficiency cut is fixed to $90\%$ for \textit{Hadronness} and $75\%$ for $\theta^2$.
Right: improvement for different SNR of the precision  in the BASiL approach relative to the standard one. The definition of precision  and bias can be found in Eq. \eqref{Eq:Res_def} and \eqref{Eq:Bias_def}, respectively. Observations are simulated assuming $\alpha b = 100$, with $\alpha = 1/3$. }
\label{Fig:ResBias_SNR}
\end{figure*}

In the previous section we fixed the SNR to $10\%$ and let the efficiency cut vary. We now want to do the opposite, i.e. study the precision  and bias by varying the SNR. For this purpose we fixed the efficiency cut in \textit{Hadronness} and $\theta^2$ to $90\%$ and $75\%$, respectively, being these values the recommended ones~\cite{Performance_Paper}  and the ones that, as one can see in Fig. \ref{Fig:ResBias_Eff}, maximize the precision  power of the standard approach.
In the BASiL approach (in which $\epsilon=1$) this time both \textit{Hadronness} and $\theta^2$ will be considered when computing the single-event likelihoods of being a signal or a background event. 
Fig.~\ref{Fig:ResBias_SNR} displays the precision  and bias for different values of SNR. As expected, in both approaches both values get worse as we decrease the signal (in the MC simulations the background is kept fixed to $\alpha b = 100$). Such worsening is, however, less pronounced in the new approach, where the precision  is about $20\%$ better for a SNR of $1\%$. If the strength of the signal is instead equal to the background noise, i.e. SNR = $100\%$, then the improvement of the BASiL method relative to the standard one is $\sim 13\%$ (see right plot of Fig. \ref{Fig:ResBias_SNR}).
One can also notice that at low values of SNR the bias increases: this is due to the fact that for weak signal rates estimates $\hat{s}$  that are close or equal to zero\footnote{Recall that for a fair comparison between the two approaches, in the standard approach $\hat{s}$ is put equal to zero whenever $N_{on} - \alpha N_{off} < 0$.} becomes more frequent, and this inevitably shifts the mean value of the distribution of $( \hat{s} - s \cdot \epsilon )/(s\cdot \epsilon)$ through positive values.

We conclude that the BASiL method is capable of estimating the signal rate more precisely, without having to select data. This is due to the introduction of the combinatorial term defined in Eq.~\eqref{Eq:combinatorialTerm}, which takes into account the likelihood of each event of being a signal or background event.

\subsection{A Spectral Energy Distribution}
\label{Subsec:CrabNebula-SED}

After having evaluated the performance of the method by using MC simulations of events observed by the MAGIC telescopes, we now apply the method on a real data set.  For this purpose we used the data\footnote{The corresponding data in FITS format are publicly available in \href{https://github.com/open-gamma-ray-astro/joint-crab/tree/master/data/magic}{https://github.com/open-gamma-ray-astro/joint-crab/tree/master/data/magic}
} 
released by the MAGIC collaboration,
 which includes 40 minutes
of Crab nebula observations chosen from the sample
used for the performance evaluation in Ref. \cite{Performance_Paper}.
This data set includes only events recorded at low zenith angles ($< 30^{\circ}$) and  under good  atmospheric conditions.
All data were taken in the wobble mode with the standard offset of $0.4^{\circ}$. Off counts were obtained using three simultaneous Off regions within the same field of view and with the same offset from the camera center as the On region. Overall effective observation time is 39.2 minutes. 

\begin{figure}
  \includegraphics[width=0.99\linewidth]{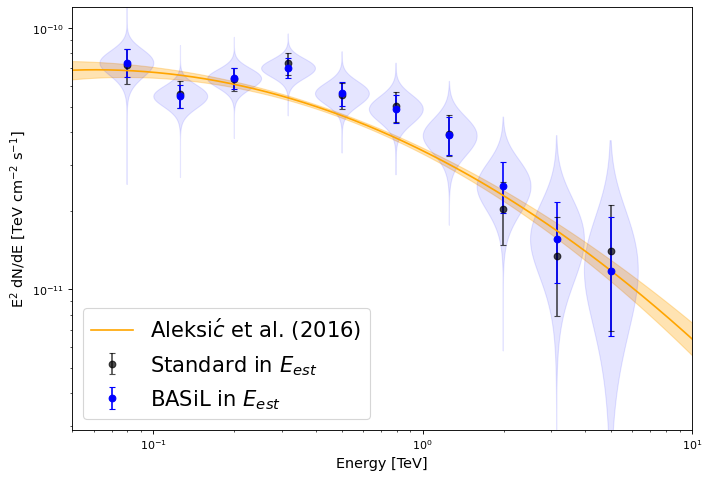} \\
    \includegraphics[width=0.97\linewidth]{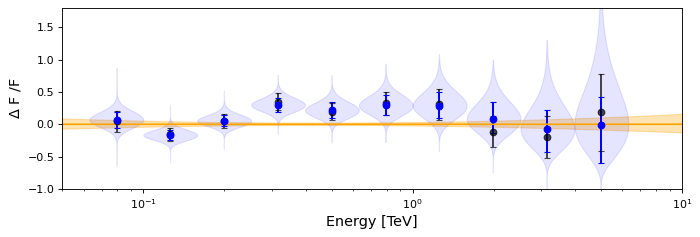} 
\caption{SED in estimated energy of the Crab Nebula (in blue) obtained by processing 0.66 hours of data released by the MAGIC collaboration with the BASiL method. For comparison also the results (in black) obtained from the same data sample using the standard analysis procedure are reported in which efficiency cuts are applied.  The blue bars of the data points are the credible interval obtained from Eq.~\eqref{Eq:CredibleInterval}, in which also uncertainties from the exposure are taken into account. ``Violin'' plots around each blue point represent the flux PDF. The obtained results are also compared to the Crab Nebula SED (in orange) from Ref. \cite{Performance_Paper} in terms of the relative flux difference.}
\label{Fig:Crab_SED_ff}
\end{figure}

The standard data analysis (whose results are shown in black in Fig.~\ref{Fig:Crab_SED_ff}) has been performed using the MAGIC Analysis and Reconstruction Software (MARS) \cite{2013ICRC...33.2937Z} where a \textit{Hadronness} and $\theta^2$
cut according to a high $\gamma$-ray
efficiency ($90\%$ and $75\%$ respectively) is applied. For the BASiL analysis instead no cut is applied on the data set. Only a global $\theta^2 < 0.08 \rm{ deg}^2$ is considered to define four identical non-overlapping regions from the center of the camera: one for the On region and three for the Off regions. 
The resulting signal rate per energy bin is reported in Tab. \ref{tab:Excess_per_EnBin}. 
It is worth comparing the signal estimation by using the BASiL approach (last column of Tab. \ref{tab:Excess_per_EnBin}) with the one we would have obtained by simply performing the difference between the total counts in the On and Off region, i.e. $N_{on} - \alpha N_{off} $ (fourth column of Tab. \ref{tab:Excess_per_EnBin}). One can notice that the BASiL approach manages to decrease by half the uncertainty in the signal estimation. This is in agreement with the result reported in Fig.~\ref{Fig:ResBias_Eff}, where the precision  of the BASiL method ($\sim 20 \% $) is about half the one obtained in the standard approach by not cutting data in \textit{Hadronness} or $\theta^2$ ($\sim 40 \% $).  

Combined with the exposure of the telescopes the values in the last column of Tab. \ref{tab:Excess_per_EnBin}  are then used to compute the spectral energy distribution (SED) points in Fig. \ref{Fig:Crab_SED_ff}.
An advantage of the BASiL approach when estimating the source flux is its capability of providing a PDF contour plot associated to each energy bin (see Fig. \ref{Fig:Crab_SED_ff}). In this way not only error bars for each flux point are drawn, but a full PDF (corresponding to the PDF in Eq. \eqref{Eq:SignalPDFTotal}) is visualized, which encodes all information we have regarding the signal estimation for that energy bin.

\begin{figure}
    \includegraphics[width=0.95\linewidth]{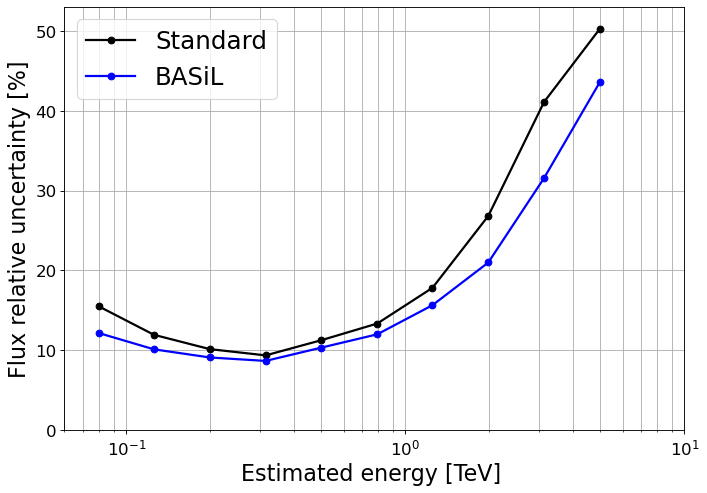} 
\caption{ Relative uncertainty in the flux estimation from Fig. \ref{Fig:Crab_SED_ff} for the standard (black) and BASiL (blue) approach.}
\label{Fig:Crab_RelError_Flux}
\end{figure}

In Fig. \ref{Fig:Crab_RelError_Flux}  we report the relative uncertainties in the flux estimation using the standard (in black) and BASiL (in blue) approach. The former is computed from
\begin{align}
    \frac{\sqrt{N_{on} + \alpha^2 N_{off}}}{ N_{on} - \alpha N_{off} } ,
    \label{Eq:Rel_unc_St}
\end{align}
with $N_{on}$ and $N_{off}$ the number of events surviving the $\theta^2$ and \textit{Hadronness} cuts in the On and Off region, respectively. 
The latter instead is obtained from
\begin{align}
    \frac{(s_{right} - s_{left})/2}{ s^*  } ,
    \label{Eq:Rel_unc_New}
\end{align}
where $s^*$ is the mode of the signal PDF and $s_{right}$, $s_{left}$ are defined in Eq. \eqref{Eq:CredibleInterval}. Uncertainties due to the exposure computation are added in quadrature, although they are negligible relative to the uncertainties in Eqs. \eqref{Eq:Rel_unc_St} and   \eqref{Eq:Rel_unc_New}.
As one can see from Fig. \ref{Fig:Crab_RelError_Flux}, relative uncertainties in the flux estimation are smaller in the BASiL approach, especially at higher energies where the signal rate is weaker. In light of the analysis performed in the previous sections using MC simulations, this is totally expected and confirms our conclusion. 

\begin{table}
\caption{\label{tab:Excess_per_EnBin} Estimated signal rate in each energy bin used for computing the  Crab nebula SED reported in Fig. \ref{Fig:Crab_SED_ff}. The signal rate in the last column is estimated from the mode of the signal PDF defined in Eq. \eqref{Eq:SignalPDFTotal}, while its uncertainties are computed using the credible interval defined in Eq. \eqref{Eq:CredibleInterval}. From the second to the fourth column  we report the counts $N_{on}$ and $N_{off}$ in the On and Off region, respectively, along with their difference $N_{on} - \alpha N_{off} \pm \sqrt{N_{on} + \alpha^2 N_{off}}$, i.e. the signal rate estimation in the frequentist approach.}
\begin{ruledtabular}
\begin{tabular}{l|lll|l}
   E [GeV] & $N_{on}$ & $N_{off}$ & Signal (freq.) & Signal (BASiL)\\  \hline

63-100 &	1714&	4494&	$216 \pm 47$&	    $204\pm23$\\[0.1cm]
100-158 &	933&	2349&	$150 \pm 35$&	    $187\pm18$\\[0.1cm]
158-251 &	622&	1327&	$180 \pm 28$   &     $185\pm16$   \\[0.1cm]
251-398 &	439&	846&	$157 \pm 23$    &     $174\pm14$   \\[0.1cm]
398-631 &	335&	593&	$137 \pm  20$   &     $114\pm11$   \\[0.1cm]
631-1000 &	215&	435&	$70 \pm 16$    &    $77.7^{+9.6}_{-8.9}$   \\[0.1cm]
1000-1585 &	132&	256&	$47 \pm 13$     &   $41.8^{+7.0}_{-6.3}$    \\[0.1cm]
1585-2512 & 	95&	203&	$27 \pm 11$      &   $21.6^{+5.1}_{-4.5}$    \\[0.1cm]
2512-3981 &	56&	140&	$9.3  \pm 8.5$    &   $8.6^{+3.4}_{-2.7}$    \\[0.1cm]
3981-6310 &	30&	83&	$2.3 \pm 6.3$      &   $3.9^{+2.3}_{-1.6} $   \\

\end{tabular}
\end{ruledtabular}
\end{table}

\section{Conclusion and outlook}
In this paper we introduced a novel method for estimating the signal rate in experiments with imprecisely measured background. Common examples are astronomical measurements at high energies, in which messengers (e.g. $\gamma$ rays, neutrinos) are detected on event by event basis, but can equally successfully be applied in particle experiments. The BASiL method, as we dubbed it, relies on the Bayesian, rather than the, more common, frequentist approach. Its main feature is that it weights events according to their individual likelihood of being signal or background, considering all the information available. 
This weighting is best summarized by the PMF of the number of signal events in Eq.~\eqref{Eq:Nexc_PDF_CombTerm}, in which the novelty of the method, i.e. the combinatorial term defined in Eq.~\eqref{Eq:combinatorialTerm}, shows up.
By doing so, BASiL avoids cutting data according to some (or a combination of) variable to suppress the background, which inevitably discards a part of the signal.
Moreover, the new method, while yielding results consistent with the standard data analysis method (see Sec.~\ref{Subsec:CrabNebula-SED} for a comparison on the example of Crab Nebula), it improves the precision   of the signal estimation, as demonstrated in Fig.~\ref{Fig:Crab_RelError_Flux}. A convenient additional feature is a PDF contour associated with each individual flux point. 
The improvement is particularly noteworthy in cases of  small signal rates (see Sec.~\ref{Sec:IACT_simulations}).  Therefore, we expect BASiL to be especially useful for analysis of data from measurements of short transients or weak signals.
Furthermore, certain investigations, such as searches for dark matter (see, e.g. \cite{2018JCAP...03..009A, 2018PDU....22...38A, 2020PDU....2800529A}), or signatures of Lorentz invariance violation (LIV, see, e.g. \cite{2008PhLB..668..253M, 2009APh....31..226M, 2017ApJS..232....9M, 2020PhRvL.125b1301A}), base their analyses on characteristics of individual events (e.g. energy, detection time). 
Depending on the values of these individual characteristics, each event contributes differently to the sensitivity of the analysis. E.g. in LIV searches, higher energy events contribute more to the sensitivity than events of lower energies. Standard data analysis methods, which rely on cuts to suppress the background, inevitably cut some signal events from the data sample, quite possibly the ones that would have contributed to the analysis sensitivity the most.
Incorporating BASiL into analysis methods could be achieved by folding each event's contribution with its likelihood of being a signal or background event. In this way, every single event would contribute with a certain weight, increasing the analysis sensitivity. At the same time, the weights would ensure that the gain in sensitivity was not artificially created.

\begin{acknowledgments}
 We would like to thank the MAGIC Collaboration for permitting the use of proprietary Monte Carlo simulations and astronomical data. We particularly would like to thank \ A.~Moralejo and J.~Sitarek for useful discussions on this method.
 
 G.D. acknowledges funding from the Research Council of Norway, project number 301718.
 
 M.D. acknowledges funding from Italian Ministry of Education, University and Research (MIUR) through the ``Dipartimenti di eccellenza'' project Science of the Universe.
 
 T.T. and J.S. acknowledge funding from the University of Rijeka, project number 13.12.1.3.02. T.T. also acknowledges funding from the Croatian Science Foundation (HrZZ), project number IP-2016-06-9782.
\end{acknowledgments}

\appendix

\begin{widetext}

\section{Derivation of equation \ref{eq:LikelihoodH1_2} }
\label{app:Derivation}

Below one finds the derivation of the
Eq.~\eqref{eq:LikelihoodH1_2}:

\begin{align*}
  p( \vec{\mathbf{x}}, \, & N_{on}, N_{off} \; | \;  s,b ; \alpha)  = 
  p(\vec{\mathbf{x}} \; | \;  N_{on},   s, \alpha b) \cdot p(N_{on}  \; | \;    s, \alpha b) \cdot p(  N_{off} \; | \;  b) \\
  &=  \prod_{i=1}^{N_{on}} \left( p(\mathbf{x}_i \, | \, \gamma) \frac{s}{s+\alpha b} +  p(\mathbf{x}_i \, | \, \bar{\gamma} ) \frac{\alpha b }{s+\alpha b} 	\right) \cdot \frac{(s + \alpha b)^{N_{on}}}{N_{on}!} e^{- s -\alpha b } \cdot \frac{ b^{N_{off}}}{N_{off}!} e^{-b}   \\
  &= \sum_{N_s = 0}^{N_{on}} \; \sum_{A \in F_{N_s}} \prod_{i \in A}  p(\mathbf{x}_i | \gamma ) \cdot \prod_{j \in A^{c} }  p(\mathbf{x}_j | \bar{\gamma} )  \, \cdot  \frac{s^{N_s} (\alpha b)^{N_{on} -N_s}}{(s+\alpha b)^{N_{on}}} \cdot \frac{(s + \alpha b)^{N_{on}}}{N_{on}!} e^{- s -\alpha b } \cdot \frac{ b^{N_{off}}}{N_{off}!} e^{-b} \\
  &= \sum_{N_s = 0}^{N_{on}} \; C(\vec{\mathbf{x}}, N_s )  \cdot \frac{s^{N_s} (\alpha b)^{N_{on} -N_s}}{N_{on}!} e^{- s -\alpha b } \cdot \frac{ b^{N_{off}}}{N_{off}!} e^{-b} \\
  &= \frac{ \alpha^{N_{on}}/N_{off}! }{ (1+\alpha)^{ N_{on} + N_{off} }}   \sum_{N_s = 0}^{N_{on}}  \frac{ (N_{on} + N_{off} - N_s)! }{(1+1/\alpha)^{-N_s}} \;  \frac{ C(\vec{\mathbf{x}}, N_s )}{N_{on}!/N_s!}  
\cdot 
 \frac{s^{N_s}}{N_s !} e^{-s} \, \cdot \, \frac{(b(1+\alpha))^{N_{on} + N_{off} - N_s} }{ (N_{on} + N_{off} - N_s)!} e^{-b(1+\alpha)} \\
 & \propto  \sum_{N_s = 0}^{N_{on}}  \frac{ (N_{on} + N_{off} - N_s)! }{(N_{on}  - N_s)! (1+1/\alpha)^{-N_s} } \;   \frac{ C(\vec{\mathbf{x}}, N_s ) }{ \binom{N_{on}}{N_s}  } \,
 \cdot \, \frac{s^{N_s}}{N_s !} e^{-s} \, \cdot \, \frac{(b(1+\alpha))^{N_{on} + N_{off} - N_s} }{ (N_{on} + N_{off} - N_s)!} e^{-b(1+\alpha)} \; .
\end{align*}
\end{widetext}

\section{A general algorithm for computing the combinatorial term}
\label{app:Algorithm}

In Sec. \ref{Sec:Single-events-Obs}  we introduced the combinatorial term $C$ defined in Eq.~\eqref{Eq:combinatorialTerm}, providing an example on how to compute it when we have $N_{on} = 3$ events in our On sample. We now want to show how the combinatorial term can be computed in a more general case without limiting ourselves to small count numbers. Let us assume we know the likelihoods of being a signal and a background event for each event $i$ with $i = 1, 2, \dots, N_{on}$. The list of likelihoods 
\begin{align*}
& p(x_1 | \bar{\gamma} ), \, p(x_2 | \bar{\gamma} ), \, \dots \, , \, p(x_{N_{on}} | \bar{\gamma} ),
\end{align*}
and
\begin{align*}
& p(x_1 | \gamma ), \, p(x_2 | \gamma ), \, \dots \, , \, p(x_{N_{on}} | \gamma ),
\end{align*}
can be respectively saved in two arrays. So that in one array we have all background likelihoods and in the other only signal likelihoods. It is important that the event order in both arrays must be the same. At this point an algorithm that takes as input these two arrays and provides  on output the combinatorial term can be easily written as follows: 


\begin{lstlisting}
algorithm Combinatorial_term
  input: array1 of length n, 
         array2 of length n
  output: array C of length n+1
  
  n <- length of array
  C[n+1] <- [1, 0, ..., 0]
  FOR i=0 to n:
    D[n] <- [0, C[0], ..., C[n-1]]
    C <- array1[i] * C + array2[i] * D
  RETURN C
\end{lstlisting}

Here \texttt{array1,array2} have to be thought as the array containing the list of background and signal likelihoods, respectively. For instance, $C(\vec{\mathbf{x}}, 2 )$ can be found in the third element of the array obtained in output
from the algorithm above defined.
Note that it may be useful when dealing with large count numbers to work with the logarithmic values of the likelihoods.

\section{Performance on different regions of the energy spectrum}
\label{app:DifferentEnergyBins}
We show in this appendix the same analysis performed and described in Sec. \ref{Sec:IACT_simulations}, but considering events simulated at lower and higher energy ranges. We will focus in particular on the same energy bins used in Ref. \cite{Performance_Paper}, namely 75-119 GeV and 754-1194 GeV. The most important feature that emerges by considering these two lower and higher energy bins is the fact that the signal/background separation better performs at higher energies. 
Such difference between low and high energy bins is caused by the fact that in IACTs the higher the energy of an event the larger and better its camera image will be.

\begin{figure*}
  \includegraphics[width=0.43\linewidth]{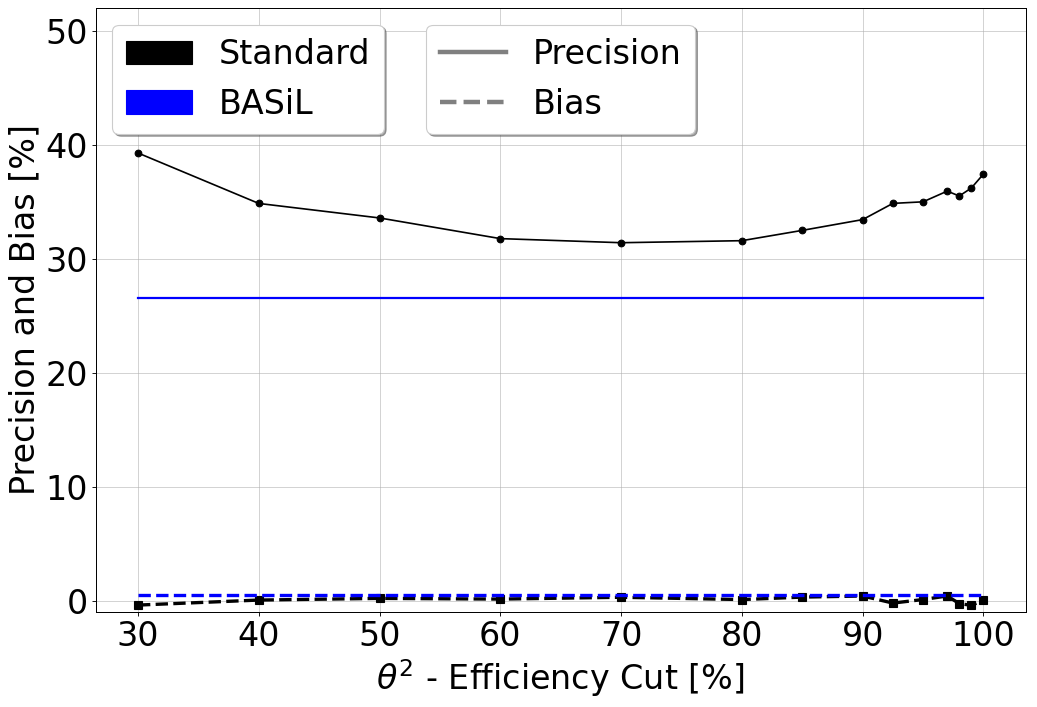} 
    \includegraphics[width=0.43\linewidth]{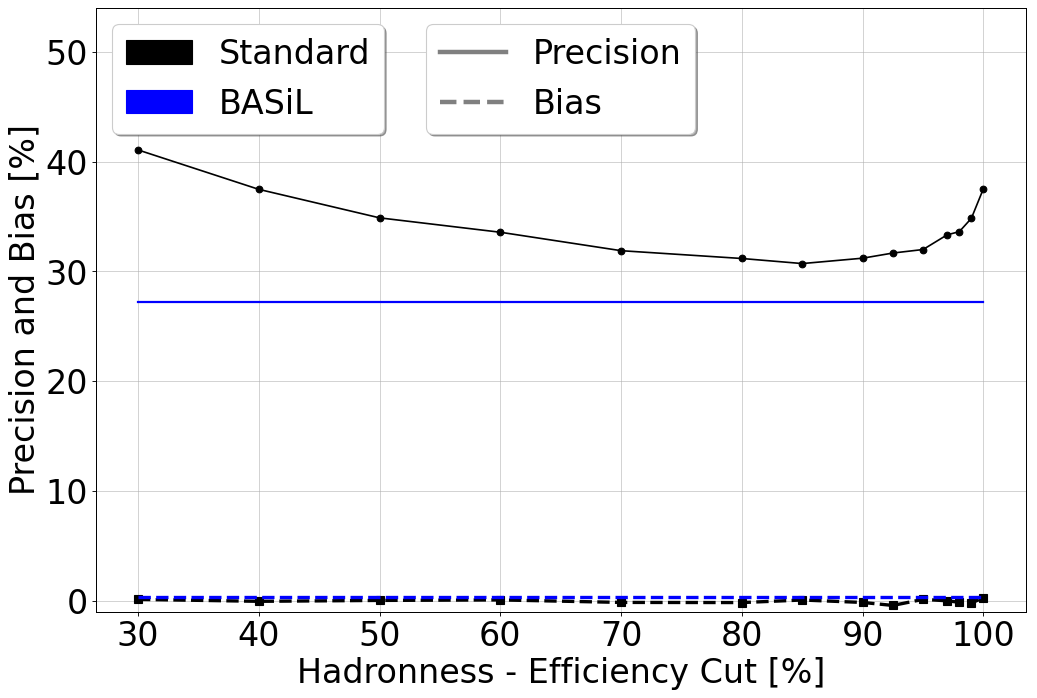} \\
    \includegraphics[width=0.43\linewidth]{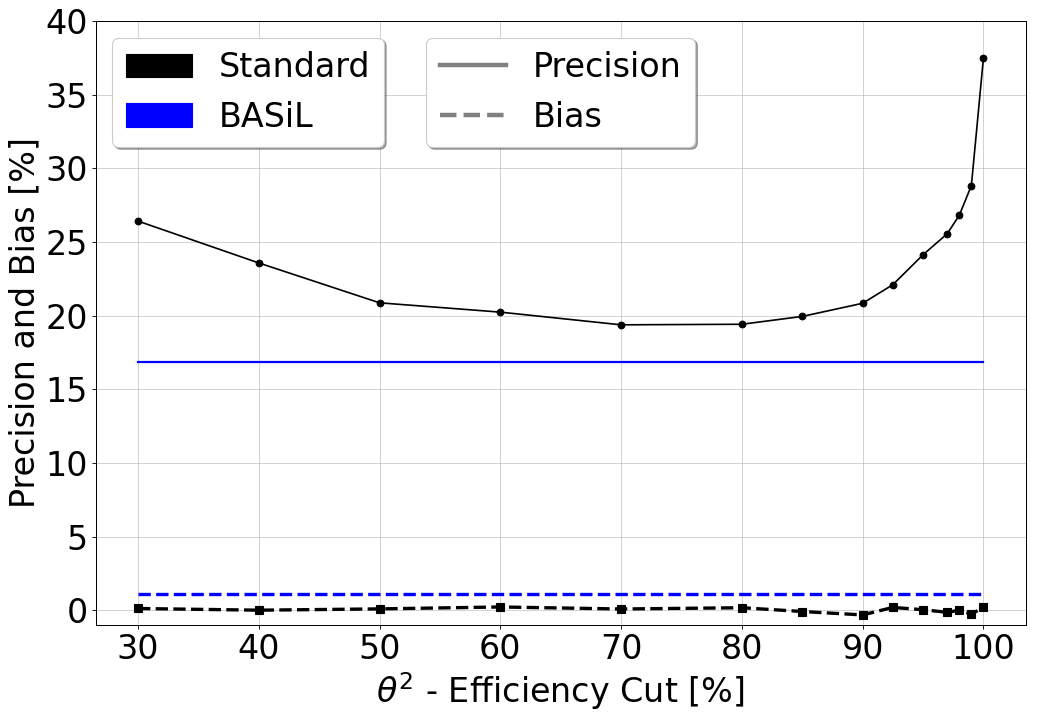} 
    \includegraphics[width=0.43\linewidth]{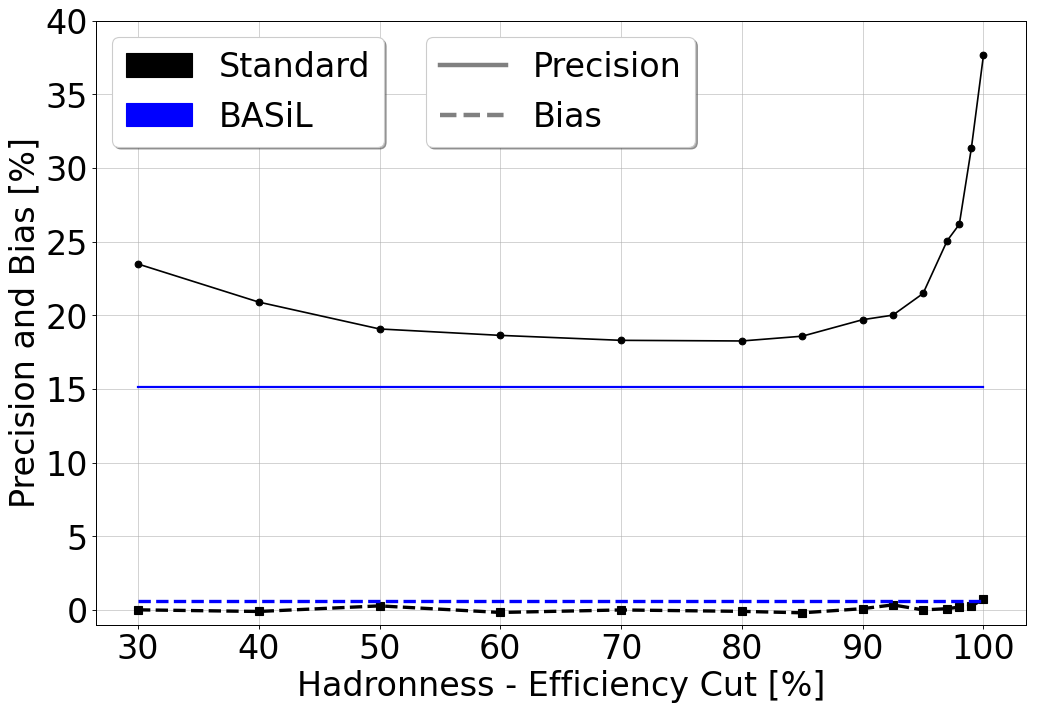}
\caption{
Comparison between the standard (black) and BASiL (blue) approach for the evolution of the precision (full line) and bias (dashed line) in the signal-rate estimation assuming different gamma-ray efficiency cut in $\theta^2$ (left) and \textit{Hadronness} (right). Note that for the BASiL approach the precision and bias do not depend on the efficiency, being $\epsilon =1$ in such case. Nonetheless, for a visual comparison these values are shown as horizontal lines.} Energy ranges considered are 75-119 GeV (top) and 754-1194 GeV (bottom).  Observations are simulated assuming $s=10^2$ and $\alpha b = 10^3$, with $\alpha = 1/3$. 
\label{Fig:ResBias_Eff_LowHighEnergy}
\end{figure*}

In Fig. \ref{Fig:ResBias_Eff_LowHighEnergy} we report the precision  and bias for different efficiency cuts applied in $\theta^2$ and \textit{Hadronness}. 
Similar conclusions made in Sec. \ref{Sec:IACT_simulations} for the medium energy bin also apply here: (i) the BASiL approach is capable of improving the signal-estimation precision  by $\sim 15 \%$ in both energy bins, (ii) the bias is always below the precision  and close to zero. 
It is also worth noticing that, as expected, the precision  increases as we go higher in energy.

\begin{figure*}
  \includegraphics[width=0.43\linewidth]{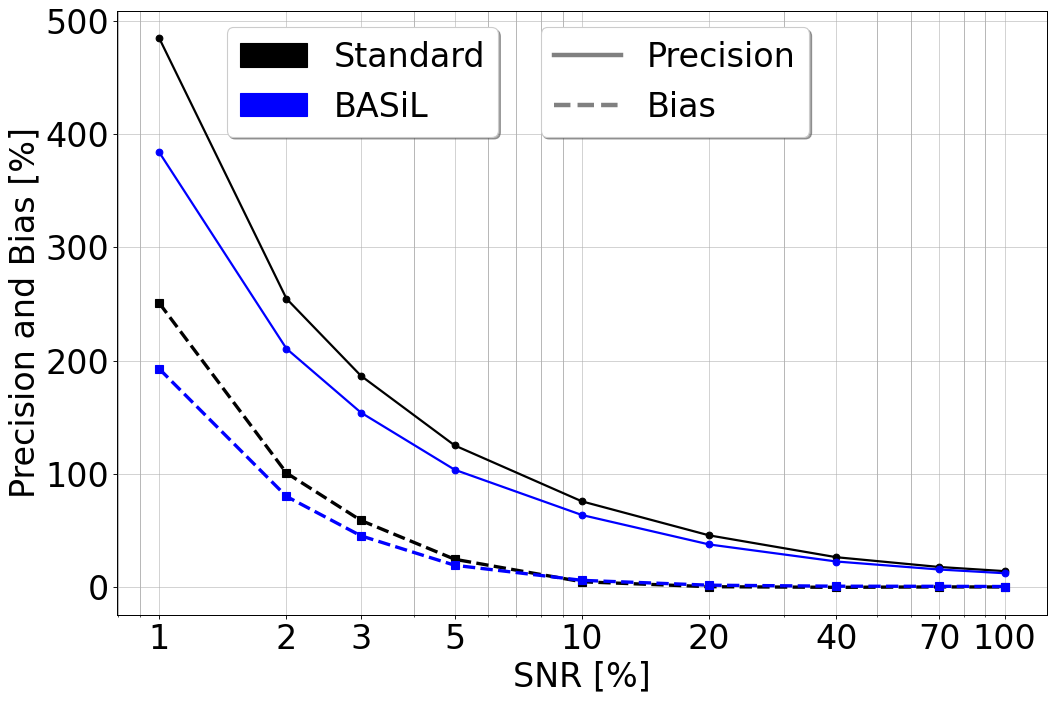} 
    \includegraphics[width=0.42\linewidth]{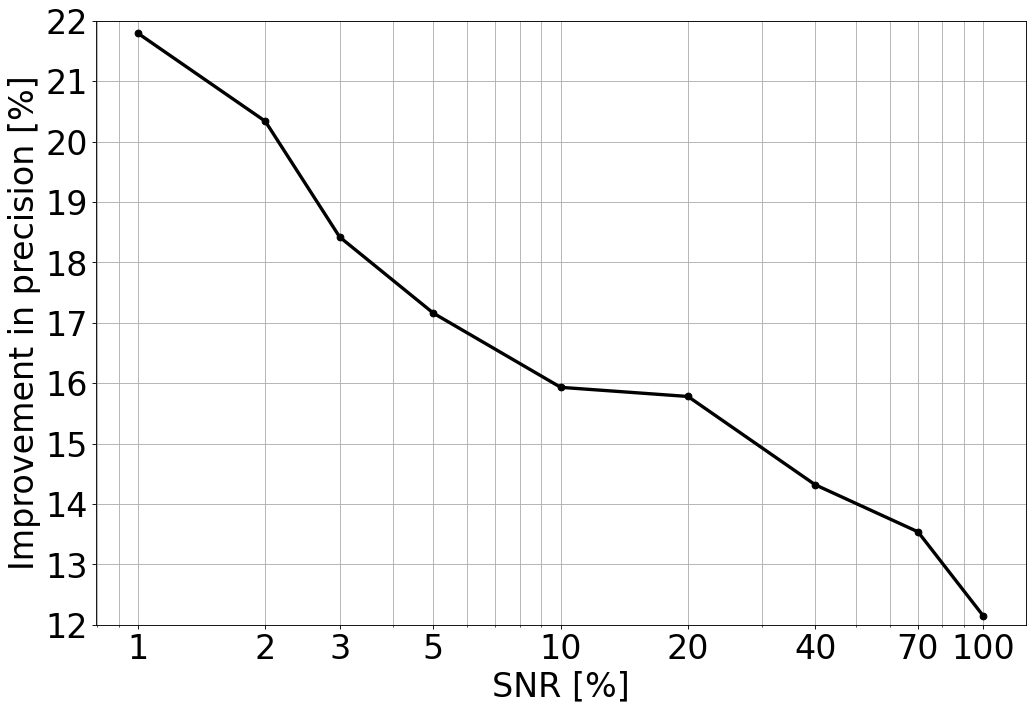} \\
    \includegraphics[width=0.43\linewidth]{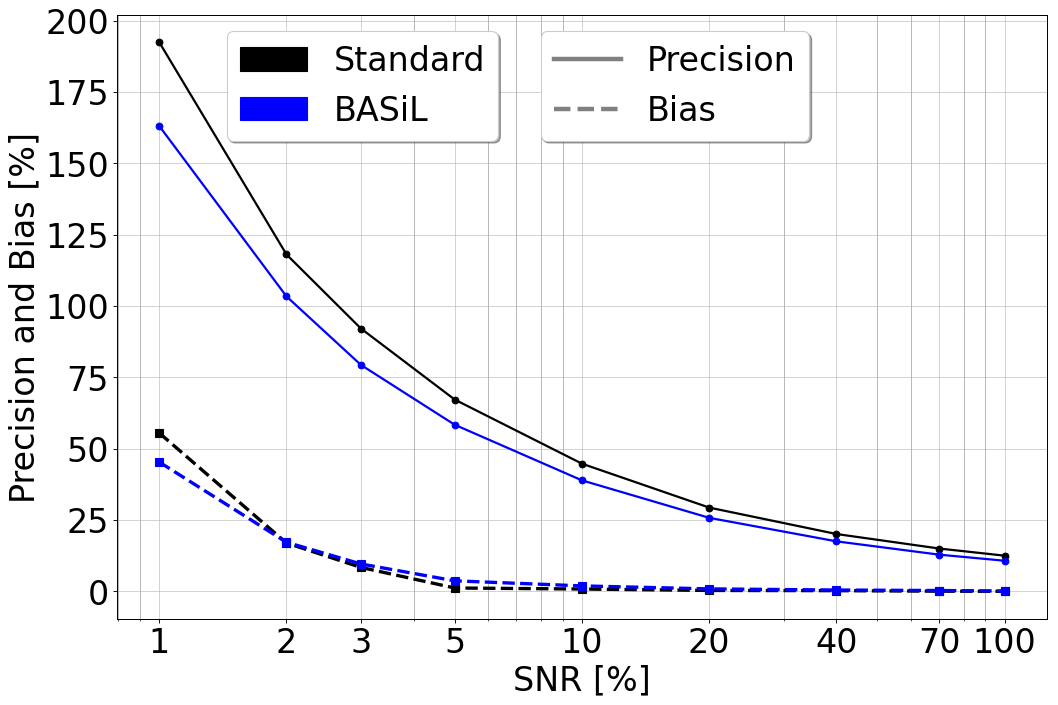} 
    \includegraphics[width=0.43\linewidth]{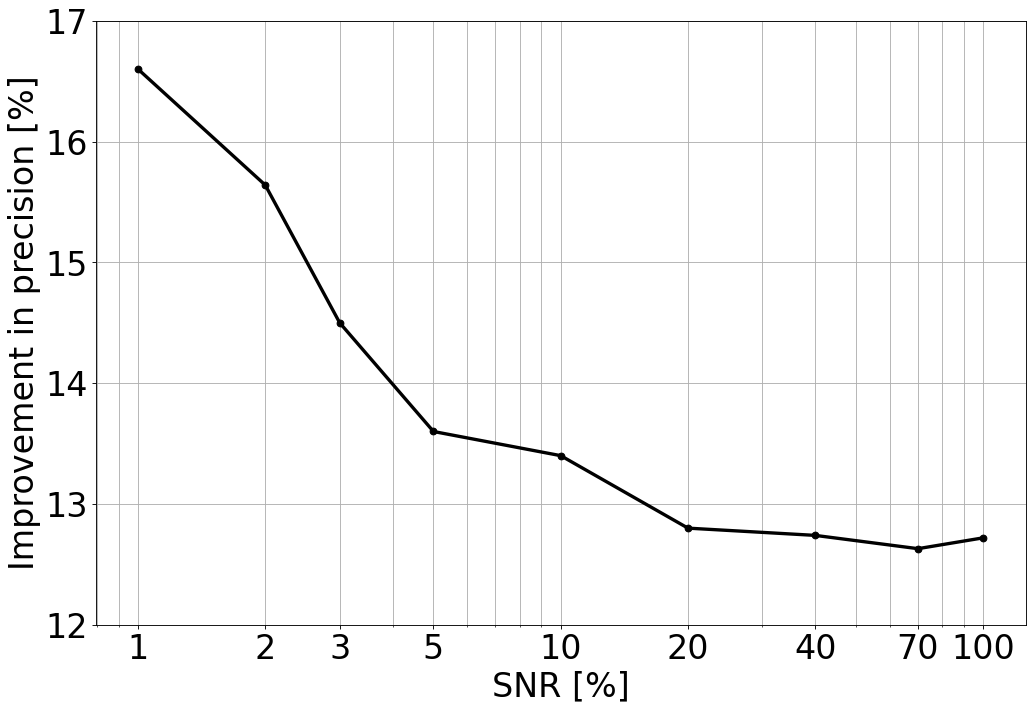}
\caption{Left: comparison between the standard (black) and BASiL (blue) approach for the evolution of the precision  (full line) and bias (dashed line) assuming different SNR. In the standard approach the efficiency cut is fixed to $90\%$ for \textit{Hadronness} and $75\%$ for $\theta^2$.
Right: improvement for different SNR of the precision  in the BASiL approach relative to the standard one. Energy ranges considered are 75-119 GeV (top) and 754-1194 GeV (bottom). Observations are simulated assuming $\alpha b = 100$, with $\alpha = 1/3$. }
\label{Fig:ResBias_SNR_LowHighEnergy}
\end{figure*}

Finally the study on the signal-estimation precision  and bias for different SNR is reported in Fig. \ref{Fig:ResBias_SNR_LowHighEnergy} where again similar conclusions of Sec. \ref{Sec:IACT_simulations} apply. 
The improvement in the precision  of the new approach increases as the SNR becomes smaller. Such improvement is  more  pronounced in the lower  energy bin (compare the upper and bottom right plots of Fig. \ref{Fig:ResBias_SNR_LowHighEnergy}).  This is due to the fact that in the higher energy bin the distinction between signal and background is more accurate and therefore signal-extraction cuts allow to remove almost all of the background while losing very few signal events.  

We conclude that the BASiL approach is capable of increasing the precision  of the signal estimation in all energy ranges and that such improvement becomes more important when the observations are background dominated.

\section{Effect of the mismatch between MC and real signal events}
\label{app:MC_real_mismatch}
\begin{figure*}[h!t]
  \includegraphics[width=0.45\linewidth]{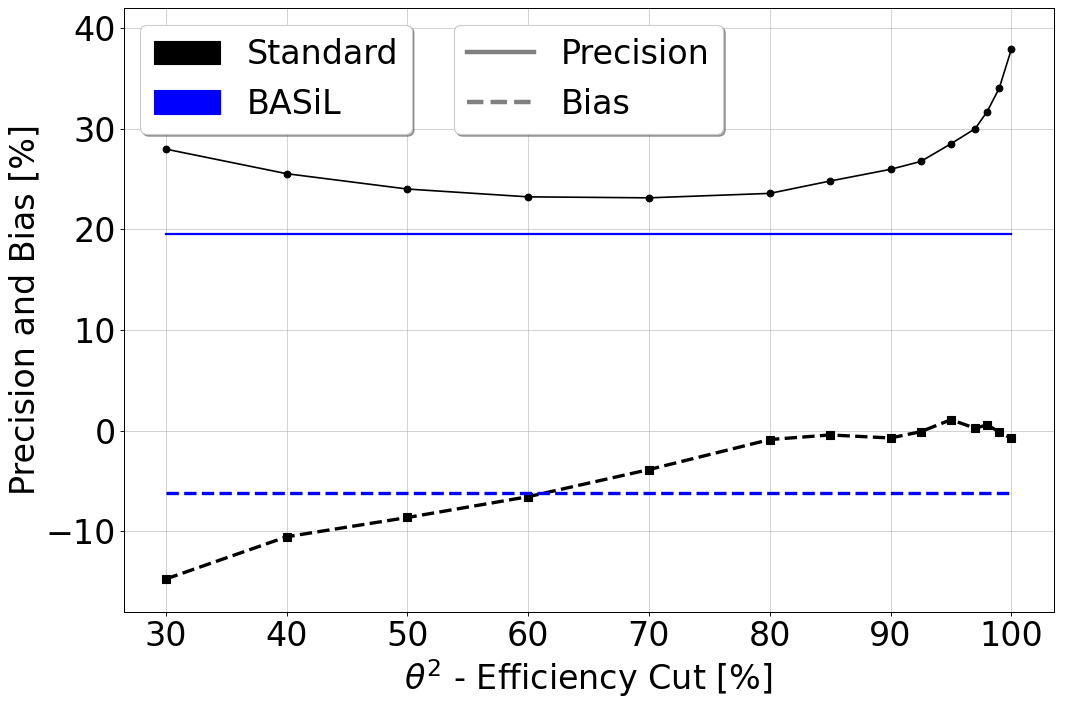} 
    \includegraphics[width=0.45\linewidth]{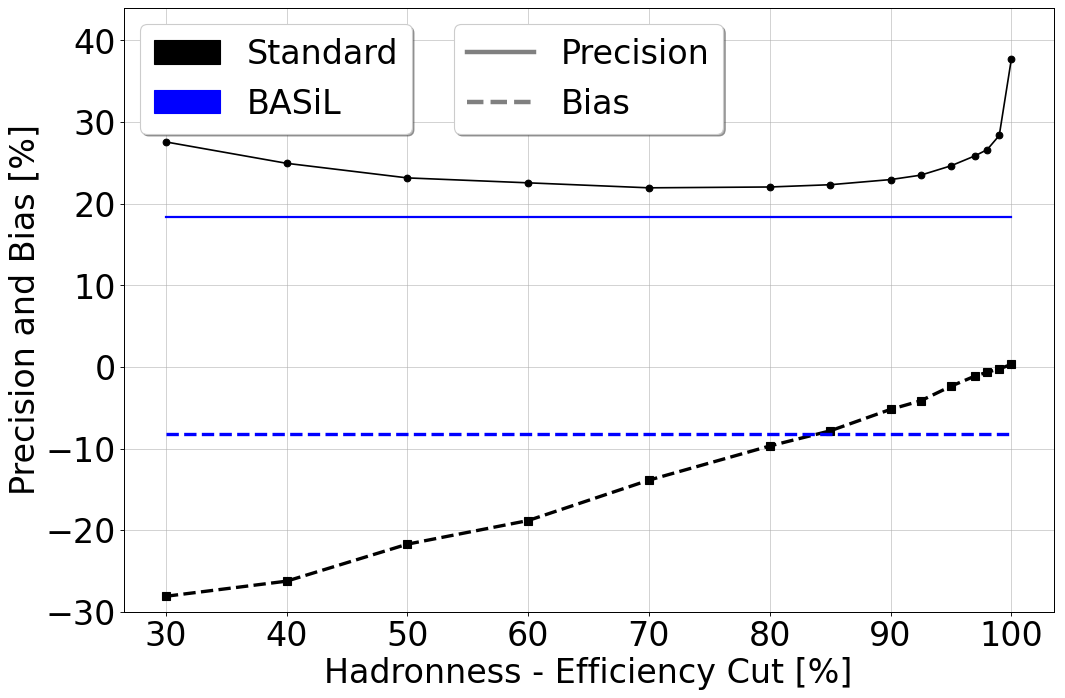} 
\caption{Same as Fig.~\ref{Fig:ResBias_Eff} but with the inclusion of the mismatch between MC and ``real'' signal population, respectively blue and red histograms of Fig. \ref{Fig:Lkl_theta2_h}}
\label{Fig:ResBias_Eff_Mismatch}
\end{figure*}

In Sec. \ref{Sec:IACT_simulations} we have studied the precision  and bias in the estimation of the signal rate by simulating On/Off measurements from the MAGIC telescopes. This estimation was performed using the standard and BASiL approach, in which the mismatch from the MC and the real gamma population  (respectively the blue and red histograms of Fig. \ref{Fig:Lkl_theta2_h}) was ignored.  We now want to study how such mismatch can affect the signal estimation. In order to do so we are going to repeat the analysis reported in Sec. \ref{Subsec:Res_bias_eff}, but this time when it comes to estimating $N_s$  only the distributions from MC-$\gamma$ events are considered: the ``real''-$\gamma$ distributions are only used in the simulation stage. The results of this analysis is reported in Fig. \ref{Fig:ResBias_Eff_Mismatch}. By comparing this figure with Fig. \ref{Fig:ResBias_Eff}, one can see that the precision  is approximately the same. The main difference, as expected, comes from the bias which is not anymore close to zero in both approaches. The bias  resulting from the mismatch between MC and signal events increases as we cut more events. 
    It is also more pronounced for the \textit{Hadronness} case:  an explanation of this can be found in Fig.~\ref{Fig:Lkl_theta2_h}, where one can see that the mismatch between MC and signal event is more pronounced for the \textit{Hadronness} case. 
 The BASiL approach produces a bias that is roughly equal to the one obtained in the standard approach by performing a cut in $\theta^2$ and \textit{Hadronness} with an efficiency around $50-60 \%$ and $70-80 \%$, respectively. It is also interesting to notice that such bias in both approaches is negative (the excess is underestimated), but anyway smaller in absolute value than the precision .
 
\bibliography{references}

\end{document}